\documentclass[12pt]{article}

\usepackage{amssymb}
\usepackage{amsmath}
\let\a=\alpha

\newcommand{\beq}{\begin{equation}}
\newcommand{\eeq}{\end{equation}}
\newcommand{\beqn}{\begin{eqnarray}}
\newcommand{\eeqn}{\end{eqnarray}}

\newcommand{\nn}{\nonumber}

\newcommand{\Tr}{{\rm Tr}}

\newcommand{\be}{\begin{equation}}
\newcommand{\ee}{\end{equation}}
\newcommand{\ba}{\begin{eqnarray}}
\newcommand{\ea}{\end{eqnarray}}
\newcommand{\bdm}{\begin{displaymath}}
\newcommand{\edm}{\end{displaymath}}

\def\a{\alpha}

\newcommand{\ie}{{\it i.e.\ }}
\newcommand{\eg}{{\it e.g.\ }}

\DeclareMathAlphabet{\mathpzc}{OT1}{pzc}{m}{it}
\def\bea{\begin{eqnarray}}
\def\eea{\end{eqnarray}}
\def\beas{\begin{eqnarray*}}
\def\eeas{\end{eqnarray*}}
\def\sla{\raise.15ex\hbox{$/$}\kern-.57em}

\def\bea{\begin{eqnarray}}
\def\eea{\end{eqnarray}}
\def\de{\partial}

\def\sla{\raise.15ex\hbox{$/$}\kern-.57em}
\def\ie{{\it i.e.}~}
\def\eg{{\it e.g.}~}
\def\ap{{\alpha^\prime}}

\def\a{\alpha}

\def\cA{{\cal A}}
\def\cB{{\cal B}}
\def\cC{{\cal C}}

\def\cE{{\cal E}}

\def\cG{{\cal G}}
\def\cH{{\cal H}}
\def\cI{{\cal I}}

\def\cK{{\cal K}}

\def\cM{{\cal M}}
\def\cN{{\cal N}}

\def\cQ{{\cal Q}}
\def\cR{{\cal R}}

\def\cT{{\cal T}}
\def\cU{{\cal U}}

\def\cW{{\cal W}}

\def\cY{{\cal Y}}
\def\cZ{{\cal Z}}

\begin{document}
\begin{titlepage}
\begin{flushright}
{ROM2F/2010/03}
\end{flushright}
\vskip 1cm
\begin{center}
{\Large\bf Pair Production of small Black Holes \\ in Heterotic String Theories}\\
\end{center}
\vskip 2cm
\begin{center}
{\bf Massimo Bianchi}, {\bf Luca Lopez} \\
{\sl Dipartimento di Fisica, Universit\`a di Roma ``Tor Vergata''\\
 I.N.F.N. Sezione di Roma ``Tor Vergata''\\
Via della Ricerca Scientifica, 00133 Roma, ITALY}\\
\end{center}
\vskip 1.0cm
\begin{center}
{\large \bf Abstract}
\end{center}
We study pair production of small BPS BH's in heterotic strings
compactified on tori and in the FHSV model. After recalling the
identification of small BH's in the perturbative BPS spectrum, we
compute the tree-level amplitudes for processes initiated by
massless vector bosons or gravitons. We then analyze the resulting
cross sections in terms of energy and angular distributions.
Finally, we briefly comment on scenari with large extra dimensions
and on generalizations of our results to non-BPS, non-extremal and
rotating BH's.



\vfill

\end{titlepage}

\section*{Introduction}

Understanding black hole physics is a challenge to any quantum
theory of gravity. The possibility that black holes be (pair)
produced in high energy collisions is a fascinating possibility
\cite{Giudice:1998ck, Giddings:2001bu, Dimopoulos:2001hw,
Giudice:2001ce, Meade:2007sz, Kanti:2008eq, Lust:2008qc,
Anchordoqui:2009mm, Gingrich:2009hj, Gal'tsov:2009zi} not without
some worries \cite{Giddings:2008gr, Giddings:2008pi}. A very
massive particle should behave as a small black hole when its
Compton length $\lambda_C = h/Mc$ is smaller than the
Schwarzschild radius $R_S = 2 G_NM/c^2$ \ie when its mass is
larger than Planck mass $M > M_{Pl} = \sqrt{\hbar c/G_N}= 1.2209
\times 10^{19} GeV/c^2$.

In perturbative string theory there is an infinite tower of very
massive states. After turning on interactions, most of them become
unstable. Some are long-lived \cite{Chialva:2004xm, Iengo:2006gm}.
Some remain stable since they are `extremal' and there is no
multi-particle state with lower mass and the same conserved
charges they can decay into. Among `extremal' states some preserve
a fraction of the original supersymmetry and are commonly called
BPS states \cite{Sen:1995cj, Dabholkar:2004yr, Dabholkar:2005dt,
Dabholkar:2007zz}.

In toroidal compactifications of the heterotic string it is very
easy to identify perturbative states of this kind
\cite{Dabholkar:1989jt, Dabholkar:1990yf, 'tHooft:1990fr,
Sen:1995in, Dabholkar:1995nc, Callan:1995hn, Horowitz:1996nw,
Damour:1999aw}. They correspond to setting the L-moving
oscillators in their (supersymmetric) ground state. KK momenta,
windings and gauge charges can be included compatibly with level
matching. At the classical level, 1/2 BPS states correspond to
charged `extremal' black-hole solutions of $\cN = 4$ supergravity
coupled to $N_v$ vector fields \cite{Sen:2009bm, Cvetic:1995kv,
Cvetic:1996kv, Youm:1997hw, Peet:2000hn}. The relevant solutions
carry only `electric' charges and display a null singularity since
their ADM mass (in the Einstein frame) vanishes at the boundary of
the $[SO(6,N_v)/SO(6)\times SO(N_v)] \times [SL(2)/O(2)]$ moduli
space. At the quantum level $R^2$ corrections can modify the
picture. Although 1/2 BPS states with a `single' charge do not
receive quantum corrections to their mass and degeneracy and thus
remain singular, 1/2 BPS states with two charges receive higher
derivative corrections \cite{Lopes Cardoso:1998wt, Sen:2004dp}
that call for Wald's entropy formula \cite{Wald:1993nt} that
generalizes Bekenstein-Hawking's entropy formula. The resulting
finite non-zero `area' of the `stretched' horizon precisely
reproduces the microscopic degeneracy of the perturbative string
spectrum \cite{Youm:1997hw, Peet:2000hn, Susskind:1993ws,
Peet:1995pe, Dabholkar:2004dq, Sen:2009bm, Prester:2010cw}. Such
states are thus good candidates for small BPS black holes whose
dynamical properties can be reliably studied in scattering
processes. Due to their charge and 1/2 BPS property, they can be
produced in pairs at least.

Our aim is to compute the cross section for pair production of
`spherically symmetric' (scalar) 1/2 BPS small BH's with
two-charges in high energy collisions of gravitons or gauge
bosons.  BPS BH's are very special in that their temperature
vanishes and they behave pretty much like very massive particles,
with a large number of degenerate microstates, accounting for
their entropy. To set the stage for a full-fledged heterotic
string computation at tree level, we first describe a field theory
toy model for pair production of charged massive scalars
\cite{Choi:1994ax}. Since the pair-produced small BH's are BPS and
stable they do not emit Hawking radiation. We analyze angular and
energy distributions of the heterotic string process with the
field theory toy model in mind and comment on scenari with Large
Extra Dimensions. We then consider a simple case with
supersymmetry broken to $\cN =2$ that enjoys particular
non-renormalization properties, the FHSV model
\cite{Ferrara:1995yx}. Finally we briefly comment on the case of
non-BPS, non `extremal' or rotating BH's. A realistic description
of macroscopic BH's with finite horizon area, even in the
classical limit, should involve wrapped branes and KK monopoles
that are dual to bound-states of D-branes in Type II or Type I
strings \cite{Maldacena:1997tm}.

The plan of the paper is as follows. In Sect. \ref{ZBPS} we recall
the partition function of perturbative 1/2 BPS states in toroidal
compactifications of heterotic strings. Moreover, we analyze three
types of 1/2 BPS states depending on their `electric' charges and
discuss the formula for their microscopic entropy. In Sect.
\ref{toymodel}, we present the field theoretic cross section for
pair production of massive charged scalars. The heterotic string
computation is described in Sect. \ref{stringprod}. We describe
both processes initiated by vector bosons and by gravitons. We
give general formulae for the amplitudes and then specialize to
the simplest non-trivial case for illustrative purposes. In Sect.
\ref{crossect} we write down explicit expressions for the cross
sections and comment on angular and energy distributions as well
as on scenari with Large Extra Dimensions. In  Sect.
\ref{BhinFHSV} we consider analogous processes in the FSHV model
\cite{Ferrara:1995yx}. We conclude in Sect. \ref{Conc} with a
summary of our results and some comments on (small) non-BPS,
non-extremal and rotating BH's.

\section{BPS partition function in Heterotic String}
\label{ZBPS}

The perturbative spectrum of heterotic strings compactified on
tori contains massless, 1/2 BPS, and long multiplets. 1/4 BPS
states with both `electric' and `magnetic' charges are
intrinsically non-perturbative in the `heterotic duality frame'
and arise from bound states of fundamental strings with K-K
monopoles or NS5-branes (`H-monopoles'). In dual descriptions, \eg
Type II on $K3\times T^2$, some 1/4 BPS states admit a
perturbative description at special points in the moduli space.
For simplicity, we will consider the component with maximal
possible rank of the gauge group ($r=6+6+16 = 28$) in the $\cN=4$
context. Rank reduction is also possible in the so-called
heterotic CHL models \cite{Chaudhuri:1995fk}, dual to Type I
models with a quantized $B$ or otherwise
\cite{Bianchi:1991eu,Bianchi:1997rf,
Witten:1997bs,Angelantonj:1999xf, Bachas:2008jv}.

In $D=10$, as a result of the GSO projection, the one-loop
partition function for heterotic strings reads
\cite{Kiritsis:2007zz}\be \cZ_{Spin(32)/Z_2} =
{\theta_3^{4}-\theta_4^{4}-\theta_2^{4} - \theta_1^{4}\over
2\eta^{12}}\times
{\bar\theta_3^{16}+\bar\theta_4^{16}+\bar\theta_2^{16}+\bar\theta_1^{16}\over
2\bar\eta^{24}} \ee and \be \cZ_{E(8)\times E(8)} =
{\theta_3^{4}-\theta_4^{4}-\theta_2^{4}-\theta_1^{4}\over
2\eta^{12}}\times
{[\bar\theta_3^{8}+\bar\theta_4^{8}+\bar\theta_2^{8}+\bar\theta_1^{8}]^2\over
4\bar\eta^{24}} \ee
 where $ \eta$ is Dedekind's function and
$\theta_\alpha$ with $\alpha =1,2,3,4$ are Jacobi functions. The
partition function vanishes thanks to Jacobi's identity, which
accounts for space-time supersymmetry in the L-moving sector.
Modular invariance results after inclusion of the bosonic and
(super)ghost zero-modes, producing a factor ${V / {Im\tau}^4}$,
that nicely combines with the modular invariant measure
$d^2\tau/Im\tau^2$. It is convenient to express $\cZ$ in terms of
the characters of the $SO(8)$ L-moving current algebra at level
$\kappa = 1$ (Little Group for massless states in $D=10$) and of
the R-moving current algebras. To this end, the characters of the
four conjugacy classes of $SO(2n)$ (vector $V$, spinor $S$,
co-spinor $C$, and singlet $O$) read \cite{Bianchi:1990tb,
Bianchi:1990yu} \be O_{2n} = {\theta_3^n + \theta_4^n \over 2
\eta^n} \: , \quad V_{2n} = {\theta_3^n - \theta_4^n \over 2
\eta^n} \: , \quad S_{2n} = {\theta_2^n + i^n \theta_1^n \over 2
\eta^n} \: , \quad C_{2n} = {\theta_2^n - i^n \theta_1^n \over 2
\eta^n}\ee Then one finds
 \be \cZ^H_{D=10} = {\cQ \bar\cG \over |\eta^{8}|^2} \ee
where $\cQ = V_8 - S_8 = ({\bf 8}_v - {\bf 8}_s) q^{1/3} +
massive$ is the super-character introduced in
\cite{Bianchi:1998vq} and $\cG = O_{32} + S_{32} = q^{-2/3} + {\bf
496} q^{1/3} + ...$ for $Spin(32)/Z_2$ or $\cG = E_8^2 \equiv
[O_{16} + S_{16}]^2 = q^{-2/3} + ({\bf 248 } + {\bf 248 }) q^{1/3}
+ ...$ for ${E(8)\times E(8)}$. As usual we set $q=e^{2\pi
i\tau}$.

\subsection{Perturbative BPS states in toroidal compactifications}

After toroidal compactification, the two heterotic strings are
continuously connected by Wilson lines breaking the gauge group to
a common sub-group of both $SO(32)$ and ${E(8)\times E(8)}$.
Setting henceforth $\ap =2$ for notational convenience, the
one-loop partition function in $D=4$ reads\footnote{While ${\bf m}
= (m_1,...,m_6)$ and ${\bf n}=(n^1,...,n^6)$ are unrestricted
6-ples of integers, the 16-uples ${\bf r}=(r_1,...,r_{16})$ belong
to an even self-dual lattice, \ie $\Gamma_{E_8}\oplus
\Gamma_{E_8}$ or $\Gamma_{Spin_{32}/Z_2}$.} \be \cZ^H_{T^6} =
\sum_{{\bf m}, {\bf n}, {\bf r}} q^{{{1\over 2}} |{\bf p}_L|^2}
\bar{q}^{{{1\over 2}} |{\bf p}_R|^2} {\cQ \over \bar\eta^{24} }\ee
where
 \be {\bf p}_L = [(E^t)^{-1}({\bf m}+
A^t{\bf r} + (B + {1\over 2} A^tA) {\bf n}) + {1\over {{2}}} E
{\bf n}; {\bf 0}] \ee are the 6 central charges of the $\cN=4$
superalgebra, the 6 graviphotons couple to, and \be {\bf p}_R =
[(E^t)^{-1}({\bf m}+ A^t{\bf r} + (B + {1\over 2} A^tA) {\bf n}) -
{1\over {{2}}} E {\bf n}; {{{}}} ({\bf r} + A {\bf n})]\ee are 22
`matter' charges, the vector bosons in $\cN=4$ vector multiplets
couple to. The moduli space $\cM = SO(6, 22)/SO(6)\times SO(22)$
is parameterized by the internal metric $G_{ij}$ or rather by the
6-bein $E^{\hat{i}}_i$, for which
$G_{ij}=\delta_{\hat{i}\hat{j}}E^{\hat{i}}_iE^{\hat{j}}_j$, the
anti-symmetric tensor $B_{ij}$ and the Wilson lines $A_i^a$ $a=1,
... 16$ \cite{Narain:1986am}.  The remaining $SL(2)/SO(2)$ is
spanned by the dilaton and axion, that belong in the $\cN =4 $
supergravity multiplet.

The level matching condition reads \be {{}} |{\bf p}_L|^2 + 2 (N_L
- \delta_L) = {{}} M^2 = {{}}|{\bf p}_R|^2 + 2 (N_R - 1) \ee where
$\delta_L$ denotes the ground-state energy in the L-moving
(supersymmetric) sector ($\delta_L^{NS} =1/2$, $\delta^R_L= 0$).
For 1/2 BPS states $N_L = \delta_L$ and one has \be 1/2 \ BPS
\quad \rightarrow \quad {\bf Q}^2 \ge -2 \ee where \be {\bf Q}^2 =
\eta_{AB} {Q}^A {Q}^B = {{}} |{\bf p}_L|^2- |{\bf p}_R|^2 = 2{\bf
n}{\bf m} - |{\bf r}|^2 = 2(N_R - 1) \ee
 is the $SO(6,22)$
invariant norm of the 28-dimensional `electric' charge vector
${\bf Q} = ({\bf n}, {\bf m}, {\bf r})$ with $\eta =
(\sigma_1\otimes {\bf 1}_{6\times 6}) \oplus (- {\bf 1}_{16\times
16})$. Perturbative states with no `magnetic' charges (${\bf P}
=0$) and ${\bf Q}^2 < -2 $ are necessarily non BPS, while states
with ${\bf Q}^2 \ge  -2 $ may be either BPS or non-BPS\footnote{We
thank Sergio Ferrara for enlightening discussions on the orbits of
$\cN =4$ states \cite{Cerchiai:2009pi}.}.

\subsection{1/2 BPS states with ${\bf Q}^2 =  -2, 0 $}

1/2 BPS states with ${\bf Q}^2 = -2 $ have $N_R = 0$ and are the
only states that can lead to gauge symmetry enhancement
\cite{Hull:1995mz, Duff:1995sm, Youm:1997hw, Peet:2000hn}. Since
${\bf Q} = ({\bf n}, {\bf m}, {\bf r}) \neq 0$, these states are
generically massive as $M^2_{BPS}= |{\bf p}_L|^2\ge 0$. For
special choices of the moduli, the conditions ${\bf p}_L= 0$ can
be satisfied giving rise to massless non-abelian vector
multiplets. In particular states with ${\bf n}{\bf m} = -1 $, \ie
${\bf r} = 0$ may become massless at self-dual points. States with
${\bf r}^2 = 2$ and ${\bf n}={\bf m} = 0 $, corresponding to the
480 `charged' vector bosons and gauginos in $D=10$, remain
massless in the absence of Wilson lines. States with ${\bf r}^2 =
2$ and ${\bf n},{\bf m}$ not all zero, still with ${\bf n} {\bf m}
= 0 $, are associated to generalized KK excitations.

1/2 BPS states with ${\bf Q}^2 =  0 $ have $N_R = 1$. For ${\bf
n}={\bf m} = {\bf r} = 0$, they correspond to the massless moduli
and their superpartners. For any other choice of ${\bf Q} = ({\bf
n}, {\bf m}, {\bf r}) \neq 0$ with ${\bf Q}^2 =  0$ one gets
massive states.

For a given set of charges $({\bf n}, {\bf m}, {\bf r})$ with
${\bf Q}^2 = -2$ ($N_R =0$) there is only one state or rather
multiplet, after including superpartners with non-zero spin, \be
d^{\cN = 4}_{1/2BPS}({\bf Q}) = 1 \quad {\rm for} \quad {\bf Q}^2
=-2\ee For ${\bf Q}^2 = 0$ ($N_R =1$) the `degeneracy' is finite:
one spin 2 multiplet ($2\times[24_B - 24_F]$ states including CPT
conjugates) and 21 spin 1 multiplets ($2\times[8_B - 8_F]$
states). We do not expect these states to correspond to smooth
classical solutions even after inclusion of higher derivative
corrections \cite{Youm:1997hw, Peet:2000hn, Dabholkar:2004dq,
Prester:2010cw}.

For a square torus without Wilson lines and antisymmetric tensor,
$G_{ij} = R^2\delta_{ij}$, $B_{ij}=0$ and $A_i^a = 0$, the $(8 +
8) \times (8 + 16 + 480)$ massless states correspond to taking the
massless ground states for both Left and Right movers \be
\cZ_{m=0} = ({\bf 8}_v - {\bf 8}_s) ({\bf 8}_v + {\bf 496}_{Adj})
= 4032_B - 4032_F\ee the minus sign accounts for the different
statistic of bosons and fermions \ie $\cZ$ is rather a Witten
index $\cI_W = tr(-)^F (q\bar{q})^H$ than a genuine partition
function.

\subsection{1/2 BPS states with ${\bf Q}^2 \ge 2 $}

1/2 BPS states with ${\bf Q}^2 \ge 2 $ are always massive inside
the moduli space, since $ M_{BPS}^2 = |{\bf p}_L|^2 = |{\bf
p}_R|^2 + {\bf Q}^2M_s^2 \ge 2M_s^2$ with $M_s = \sqrt{2/\ap}$.
BPS states with ${\bf r}\neq 0$ are `charged' wrt the `visible'
gauge group, already present in $D=10$. For fixed charges, masses
are moduli dependent. Keeping $M_{Pl}$ \ie $G_N$ fixed\footnote{In
$D=10$, one has $2\kappa_{10}^2 = (2\pi)^7 (\ap)^4
(g_s^{(10)})^2$.}, \be M_s = (2\pi)^3 g_s^{(4)}M_{Pl} \quad {\rm
where} \quad g_s^{(4)} = g_s^{(10)}\sqrt{{2\over
\hat{V}_{T^6}}}\ee with $\hat{V}_{T^6}$ the volume of the
six-dimensional internal torus in string length units $\ell_s =
\sqrt{\ap/2}$. As already mentioned $M_{BPS}$ is extremized at the
boundary of moduli space \cite{Bianchi:2009wj, Bianchi:2009mj}
where $M_{BPS}=0$ since $g_s^{(4)}\rightarrow 0$. Keeping
$g_s^{(4)}$ fixed and non-zero, the mass is extremized at points
where ${\bf p}_R = 0$ and $M_{BPS} = (2\pi)^3
g_s^{(4)}M_{Pl}\sqrt{2N_R-2}$, which can be kept hierarchically
smaller than $M_{Pl}$ for extremely small $g_s^{(4)}$.

The large degeneracy of BPS states with fixed charges is related
to the exponential growth with $N_R$ of the number of states for
the transverse R-moving bosonic oscillators. Neglecting spin, one
indeed finds \cite{Kiritsis:2007zz} \be d^{\cN = 4}_{1/2BPS}(N_R)
\approx e^{4\pi \sqrt{N_R}} \ee for $N_R >> 1$.

As mentioned in the Introduction, although the ADM mass vanishes
at the boundary of moduli space and the mass at the horizon
classically vanishes \cite{Ferrara:1995ih, Ferrara:1996dd,
Bellucci:2007ds}, including higher derivative corrections makes
the BH solutions with $ {\bf Q}^2
> 0$ become smooth and acquire a non-vanishing area that, for $
{\bf Q}^2
>> 1$ reproduces the microscopic entropy \be S_{BH} \approx 4\pi
\sqrt {{1\over 2} {\bf Q}^2}\ee resulting from the exponential
degeneracy of string states.

1/2 BPS multiplets with ${\bf Q}^2 \ge 2$ may include states with
higher spin, \ie $J>2$. For $J_{_{HWS}} = J+1$, so that
$J_{_{LWS}} = J-1$ (for $J\neq 0$),  these multiplets contain
$(2J+1)(8_B - 8_F)$ complex charged states. The multiplicities of
states with a given spin coincide with the dimensions of
representations of $Sp(4)$ that rotates the 4 real supercharges
acting as raising and as many as lowering operators. The above
degeneracy is computed for fixed mass and conserved internal
charges but without fixing the spin of the state or multiplet. One
can refine the analysis and compute the character valued partition
function that yields the degeneracy for fixed spin $J$. The result
crucially depends on whether $J \approx J_{Max} = N_R$ (at fixed
$N_R$) or not \cite{Russo:1994ev, Matsuo:2009sx}.

\section{Pair Production of Scalars in Field Theory} \label{toymodel}

For later comparison with the heterotic string results and to fix
the notation, we now briefly analyze pair production of charged
scalars $\phi$ of mass $M$ in the collisions of vector bosons and
gravitons.

Since any theory at tree level can be viewed as the truncation of
a supersymmetric theory, transition amplitudes are formally
supersymmetric \cite{Dixon:1995xx}. In particular, all 4-point
amplitudes we consider are MHV (Maximally Helicity Violating) in
the formal limit $M\rightarrow 0$. Moreover, at least at tree
level \cite{Bern:1998ug, Elvang:2007sg, Bianchi:2008pu} and in
some very special case beyond \cite{Bern:2006kd, Bern:2008pv},
(super)gravity amplitudes can be expressed as squares of
(color-ordered) gauge theory amplitudes. In turn, each term in a
gauge theory amplitude factorizes into a part which depends on the
charges or other gauge quantum numbers and a part which depends on
the spins and other kinematical variables. For fixed group theory
structure, the amplitude must be gauge invariant.

For instance, in scalar QED the amplitude for Compton scattering
or pair production/annihilation reads
\begin{equation}
\mathcal{A}_{\gamma \phi}= -
2q_e^2\left(\tilde{a}_2\tilde{a}_3\right)
\end{equation}
where
\begin{equation}
\label{aTilde} \tilde{a}_i=a_i-\frac{p_1a_i}{p_1k_i}k_i
\end{equation}
are manifestly gauge invariant combinations of the (incoming)
photon polarizations $a_i$ ($i=2,3$). For given helicity $\lambda$,
$a_\mu^{(\lambda)}(k)$ satisfies:
\begin{equation}
k\cdot a^{(\lambda)}(k)=0, \ \ a^{(\lambda)}(k)\cdot
a^{(\lambda')}(k)=\delta_{\lambda,-\lambda'},
\end{equation}
where $k_2$ and $k_3$ denote the $4$-momenta of the
photons\footnote{For compactness, we sometimes use the notation
$p_2$ and $p_3$ for $k_2$ and $k_3$, \ie in the argument of the
$\delta$ function of momentum conservation.} ($k_i^2 = 0$) and
$p_1$ and $p_4$ denote the $4$-momenta of the scalars ($p_i^2 =
M^2$).

In (super)gravity the situation seems daunting at first look. The
Lagrangian relevant for the process consists of many terms
obtained by expanding in the weak field limit around flat
Minkowski space-time\footnote{As customary in String Theory, we
use mostly plus signature. This entails various sign changes with
respect to Field Theory formulae in \eg \cite{Choi:1994ax}.
Moreover $\kappa_{ref\cite{Choi:1994ax}} = 2 \kappa_{standard}$.}
$g_{\mu\nu}=\eta_{\mu\nu}+{2}\kappa_g h_{\mu\nu}$, where $\kappa_g
=\sqrt{8\pi G_N}$, with $G_N$ Newton's constant, up to
$o\left(h^3\right)$ (linearized gravity). In addition one has to
keep the terms involving the interactions between the scalar field
($\phi$) and the graviton ($h_{\mu\nu}$). Quite remarkably the
gravitational scattering amplitude is essentially the square of
the scattering amplitude in scalar QED \cite{Choi:1994ax}. We will
find the same result in the string theory approach. This should
not come as a surprise in view of the KLT relations between open
and closed string amplitudes \cite{Kawai:1985xq}.

Either by brute force computation of the relevant Feynmann
diagrams or by exploiting KLT-like relations with gauge theory
amplitudes, the resulting transition amplitude reads
\cite{Choi:1994ax}
\begin{equation}
\mathcal{M}_{h \phi}=\frac{\kappa_g^2}{2q_e^4}F\mathcal{A}_{\gamma
\phi}^2
\end{equation}
where
\begin{equation}
\label{KinFactF}
F=\frac{(p_1k_2)(p_1k_3)}{k_2k_3}
\end{equation}
is an ubiquitous kinematical factor.

 Linearized general
coordinate invariance allows the decomposition of the graviton
spin-2 polarizations $h_{\mu\nu}^{(2\lambda)}$ into a product of
two spin-1 polarizations
\begin{equation}
h_{\mu\nu}^{(2\lambda)}=  a_\mu^{(\lambda)} a_\nu^{(\lambda)} =
h_{\nu\mu}^{(2\lambda)}
\end{equation}
so that the polarization tensor $h_{\mu\nu}^{(2\lambda)}$ is
symmetric and satisfies:
\begin{equation}
k^\mu
h_{\mu\nu}^{(2\lambda)}=h_{\mu\nu}^{(2\lambda)}k^\nu=h^\mu_{\phantom{\mu}\mu}=0,
\end{equation}
with the graviton $4$-momentum $k$.

Defining \be f = {M^2 \over 2F} = - {M^2 \over p_1 k_2} - {M^2
\over p_1 k_3} \ee one finds (for Compton scattering) \be
\cM_{\lambda_1,\lambda_2}^{h\phi} = 2{\kappa_g^2} F
(\delta_{\lambda_1,\lambda_2} + f)^2 = 2{\kappa_g^2} F [(1+f)^2
\delta_{\lambda_1,\lambda_2} + f^2 \sigma_{\lambda_1,\lambda_2}]
\ee where $\delta_{\lambda_1,\lambda_2}$ is helicity preserving
and $\sigma_{\lambda_1,\lambda_2}$ is helicity flipping. Note that
for $M=0$ ($f=0$) only the helicity preserving amplitude survives.
If both gravitons are incoming, as in pair production processes,
helicity flipping and preserving amplitudes get exchanged.

In order to compute the cross section, one needs \be
|\cM_{\lambda_1,\lambda_2}^{h\phi}|^2 = 4 {\kappa_g^4} F^2 \{
[(1+f)^4 + f^4]\delta_{\lambda_1,\lambda_2} + 2 f^2(1+f)^2
\sigma_{\lambda_1,\lambda_2} \} \ee Averaging over helicities of
the incoming gravitons one gets \be
\langle|\cM_{\lambda_1,\lambda_2}^{h\phi}|^2\rangle = 2 \kappa_g^4
F^2 \{ [(1+f)^4 + f^4] + 2 f^2(1+f)^2 \} \ee

In the CM, the kinematics for incoming momenta reads \be k_2 =
(E,\vec{k}) \qquad k_3 = (E,-\vec{k}) \qquad p_1 = (-E,\vec{p})
\qquad p_4 = (-E,-\vec{p}) \ee with $k_2 + k_3 + p_1 + p_4 = 0 $,
$|\vec{k}| = E$ and $|\vec{p}|=\sqrt{E^2-M^2}$. As a result \be s
= - (k_2 + k_3)^2 = 4E^2 \quad  t = - (k_2 + p_1)^2 = M^2 -
2E^2-2\vec{k}\cdot \vec{p} \quad  u = - (k_3 + p_1)^2 = M^2 -
2E^2+2 \vec{k}\cdot \vec{p} \ee

Moreover, indicating by $\theta$ the scattering angle so that
$\vec{k}\cdot \vec{p} = |\vec{k}||\vec{p}|\cos\theta$, one gets
 \be F=\frac{(p_1k_2)(p_1k_3)}{k_2k_3} = -{1\over 2}
[E^2 \sin^2\theta + M^2\cos^2\theta] \quad , \quad f = {M^2\over 2
F} = - {M^2 \over E^2 \sin^2\theta + M^2\cos^2\theta} \ee Setting
$\eta \equiv E/M \ge 1$ (threshold for the process) one eventually
finds \bea &&{d\sigma \over d\Omega} = {\kappa_g^4 M^2\over 32
(4\pi)^2 } {\sqrt{\eta^2 - 1} \over \eta^3 [\eta^2 \sin^2\theta +
\cos^2\theta]^2} \{2 - 4 [\eta^2 \sin^2\theta + \cos^2\theta]
\\
&&+ 4 [\eta^2 \sin^2\theta + \cos^2\theta]^2 - 2 [\eta^2
\sin^2\theta + \cos^2\theta]^3 + {1\over2} [\eta^2 \sin^2\theta +
\cos^2\theta]^4\} \nn \eea

Integrating over the solid angle yields the total cross section
\begin{eqnarray}
\sigma =\frac{\pi M^2\sqrt{\eta^2-1}}{2M_{Pl}^4\eta^3}
\left[\frac{1}{\eta^2}+\frac{1-4\eta^2}{2\eta^3\sqrt{\eta^2-1}}
\log\left(\frac{\eta+\sqrt{\eta^2-1}}{\eta-\sqrt{\eta^2-1}}\right)
+\frac{4\eta^4}{15}-\frac{6\eta^2}{5}+\frac{103}{30}\right]
\end{eqnarray}
that displays the characteristic growth with the square of the
energy at large $E$, justifiable on purely dimensional grounds
\cite{Giddings:2001bu,Dimopoulos:2001hw}. If one were to interpret
the charged massive scalars as small BPS BH's of opposite charge,
even a very crude estimate for the cross-section of pair
production in (super)gravitons collisions would require the
inclusion of degeneracy factors $d_{BH} ({\bf Q}) \approx
\exp{S_{BH}({\bf Q})}$ and possibly of form factors. At CM
energies of the order of some TeV's the above process has a
vanishingly small cross-section, unless the fundamental scale of
gravity be much lower than $M^{(4)}_{Pl}$
\cite{ArkaniHamed:1998rs, Antoniadis:1998ig}. Yet similar
processes are expected to take place even at LHC after replacing
the gravitons with gluons or quarks and the massive complex
scalars with stable not necessarily BPS BH's charged wrt the
Standard Model as can appear in superstring (flux)
compactification.

In the following we will address the problem in the largely
simplified, yet tractable, context of heterotic compactifications
on tori and simple orbifolds.

\section{Pair Production Amplitudes for
Heterotic Strings} \label{stringprod}

We have previously seen that string states that correspond to
small BH's with two charges can be pair produced in graviton or
gauge boson collisions at very high energies. Let us proceed and
compute the tree-level amplitude for these processes. For
simplicity we will mostly focus on the subspace of moduli space
with zero Wilson lines, where a distinction between a `visible'
and a `hidden' gauge groups is possible. In our conventions, the
former corresponds the non abelian gauge bosons already present in
$D=10$. The latter corresponds to the mixed components of the
metric and anti-symmetric tensor with generically abelian
symmetry. At a generic point in the moduli space of toroidal
compactifications such a distinction makes little or no sense,
since the various vectors can mix with one another. It becomes
meaningful again in phenomenologically more interesting cases with
lower or no supersymmetry.

The amplitudes for charged scalar BH pair production in vector
boson or graviton collisions are given by \be \cA_{vv\rightarrow
\Phi\bar\Phi}= \langle V_{\Phi} V_{v} V_{v} V_{\bar\Phi} \rangle
\ee and \be \cM_{hh\rightarrow \Phi\bar\Phi} = \langle V_{\Phi}
V_{h} V_{h} V_{\bar\Phi} \rangle \ee where $V_v$, $V_h$ and
$V_{\Phi}$ are vertex operators for vector bosons, gravitons and
small BH's.

\subsection{Vertex operators}

Up to normalization factors, to be discussed momentarily, in the
canonical superghost picture, the gauge boson vertex operator is
\cite{Kiritsis:2007zz} \be V_{v} = a_\mu e^{-\varphi}\psi^\mu
\bar{J}^a e^{ik\cdot X} \ee with $k^2=k\cdot a = 0$, the graviton
vertex operator is \cite{Kiritsis:2007zz} \be V_{h} = h_{\mu\nu}
e^{-\varphi}\psi^\mu \bar\de X^\nu e^{ik\cdot X} \ee with
$h_{mu\nu} = h_{\nu\mu}$ and $k^2=k^\mu h_{mu\nu} = h^\mu{}_\mu
=0$, and the two-charge massive scalar vertex is \be V_{\Phi} =
\Phi^{(N_R)}_i e^{-\varphi}\psi^i e^{i{\bf p}_L{\bf X}_L} e^{i{\bf
p}_R{\bf X}_R} e^{ipX} \ee where $p^2 = -M^2 = - |{\bf p}_L|^2$
and \be \Phi^{(N_R)}_i = \Phi_{i,j_1...j_n} \bar\de^{\ell_1}
X_R^{j_1}...\bar\de^{\ell_n} X_R^{j_n} \ee is a polynomial of
degree $N_R = \sum_r \ell_r$ in the derivatives of the R-moving
bosonic coordinates. As already said, $N_R$ is determined by level
matching to be $N_R = 1 + {\bf n}  {\bf m} - {1\over 2} |{\bf
r}|^2 = 1 + {1\over 2} {\bf Q}^2$. Similar arguments apply to the
super-partners of the scalar BH's under consideration. For
simplicity we will mostly focus on spherically symmetric small
BH's, whereby $j_r$ label internal coordinates $X^i_R$ with
$i=1,..., 6$ or R-moving (non-abelian) currents $\bar{J}^a$, with
$a=1,...,{\rm dim}G$. BPS states or rather multiplets with higher
spins require inclusion of $\bar\de^\ell X^\mu_R$ and will be
briefly considered at a later stage together with non-BPS and
non-extremal BH's.

Due to super-ghost number violation at tree level, one needs the
vertex operators for vector bosons and gravitons with super-ghost
number $q=0$, that read \be V^{(0)}_{v} = a_\mu\left(\partial
X^\mu+i\left(k \cdot \psi\right)\psi^\mu\right)\bar{J}^a
e^{ik\cdot X} \ee and \be V^{(0)}_{h}=h_{\mu\nu}\left(\partial
X^\mu+i\left(k \cdot \psi\right)\psi^\mu\right)\bar{\partial}X^\nu
e^{ik\cdot X} \ee In order to get the right dependence on
$g_{_{YM}}$ and $G_N$ in the formal field theory limit $\ap
\rightarrow 0$ one has to  dress the vertex operators with the
normalization factors \be N_v = N_\Phi = g_{_{YM}} \sqrt{2\over
\ap} \quad , \quad N_{h} =\frac{4\sqrt{\pi\kappa}}{\ap M_{Pl}}
 \ee where $\kappa$ is the level of the current algebra,
$M_{Pl} = (2\pi)^{-3} (g_s^{(4)})^{-1} \sqrt{2/\ap}$ and
 \be g_{_{YM}} =
{g_s^{(4)}\over \sqrt{\kappa}} = {g_s^{(10)}} \sqrt{2\over
\kappa\hat{V}_{T^6}} \ee and include a factor \be N_{sphere} =
{(2\pi)^4\delta(\sum_i p_i) \hat{V}_{T^6} \over (g_s^{(10)})^2
(\ap/2)^2} \ee for the sphere.

\subsection{Correlation functions}

 Apart from the space-time bosonic zero-modes that implement
momentum conservation and the internal bosonic zero-modes that
produce a factor $V_{T^6}$, all world-sheet correlation factorize
into Left- and Right- movers\footnote{Overall numerical and $g_s$
dependent factors will be reinstated at the end.}
\begin{equation}
\cW (z_i, \bar{z}_i) =\cW_L (z_i)\cW_R(\bar{z}_i)
\end{equation}
Neglecting the bosonic ghosts and setting $P_{L,i} = (p_i, \pm
{\bf p}_L)$, for the L-movers one has
\begin{equation}
\cW_L (z_i) = \langle e^{-\varphi}\psi^i e^{iP_{L,1}{\bf X}_L}
a_2\cdot (\partial X - i \psi k_2\psi)e^{ik_2X_L} a_3\cdot
(\partial X - i \psi k_3\psi)e^{ik_3X_L} e^{-\varphi}\psi^j
e^{iP_{L,4}{\bf X}_L}\rangle
\end{equation}
that further factorizes as $\cW_L (z_i) = \cW^{int}_L
(z_i)\cW^{s-t}_L (z_i)$ into an internal part \be \cW^{int}_L
(z_i) = \langle e^{-\phi} \psi^i e^{i {\bf p}_L {\bf X}_L}(z_1)
e^{-\phi}\psi^je^{-i{\bf p}_L {\bf X}_L}(z_4)\rangle =
{\delta^{ij}}z_{14}^{-1-{{}}|{\bf p}_L|^2} \ee with manifest
$SO(6)$ R-symmetry and a space-time part \be \cW^{s-t}_L(z_i) =
\langle e^{ip_1 X_L}(z_1) a_2\cdot (\partial X - i \psi
k_2\cdot\psi)e^{ik_2 X_L}(z_2) a_3\cdot (\partial X - i \psi
k_3\cdot\psi)e^{ik_3X_L}(z_3) e^{ip_4{X}_L}(z_4)\rangle \ee
Setting $\cW^{s-t}_L(z_i) = \cB^{s-t}_{L}(z_i) +
\cC^{s-t}_{L}(z_i)$ one has
\bea\label{Bs-t}\cB^{s-t}_{L}(z_i)&=&\langle e^{ip_1 X_L}(z_1)
a_2\cdot\partial X e^{ik_2 X_L}(z_2) a_3\cdot
\partial X e^{ik_3 X_L}(z_3)
e^{ip_4 X_L}(z_4)\rangle \nn \\
&=&- {{}}
\left(\frac{a_{2}a_{3}}{z_{23}^2}+{{}}\sum_{r\neq2}\frac{a_2p_r}{z_{2r}}
\sum_{s\neq3}\frac{a_3p_s}{z_{3s}}\right) \mathcal{I}_L(p_i,z_i)\\
\cC^{s-t}_{L}(z_i)&=&\langle e^{ip_1X_L}(z_1) ik_2\cdot\psi
a_2\cdot\psi e^{ik_2 X_L}(z_2) ik_3\cdot\psi a_3\cdot\psi
e^{ik_3 X_L}(z_3) e^{ip_4 X_L}(z_4)\rangle \nn \\
&=& {{1\over 2}}{{}}\frac{1}{z_{23}^2} [(a_2a_3)(p_2p_3)-
(a_2p_3)(p_2a_3)] \ \mathcal{I}_L(p_i,z_i)
\end{eqnarray}
where $\cI_L(p_i,z_i)=\prod_{i<j}z_{ij}^{{{}}p_i \cdot p_j}$
is the L-mover Koba-Nielsen factor. \\

For the R-movers one has to compute correlation functions of the
form
\begin{equation}
\cW_R(\bar{z}_i)= \langle \Phi^{(1)}_{N_R} e^{i\mathbb{P}_1\cdot
\mathbb{X}_R}(\bar{z}_1)
\bar\de\mathbb{X}_R^{M}e^{i\mathbb{K}_2\cdot
\mathbb{X}_R}(\bar{z}_2)\bar\de\mathbb{X}_R^{N}e^{i\mathbb{K}_3\cdot
\mathbb{X}_R}(\bar{z}_3)\Phi^{(4)}_{N_R}
e^{i\mathbb{P}_4\cdot\mathbb{X}_R}(\bar{z}_4)\rangle
\end{equation}
where $\mathbb{P}_i=(p_i, \pm {\bf p}_{R})$ satisfy
$\mathbb{P}_R^2=p_i^2 + |{\bf p}_{R}|^2 = - M^2 + |{\bf p}_{L}|^2
- {{2}}(N_R-1)$, while $\mathbb{K}_i=(k_i, {\bf 0})$ satisfy
$\mathbb{K}_i^2 = k_i^2 = 0$. Depending on the incoming particles
the indices $M,N$ run over space-time ($\mu,..$), `hidden'
($i,...$) or `visible' ($a,...$) gauge group respectively.

Representing $\Phi_{N_R} = \Phi_{I_1...I_n} \bar\de^{\ell_1}
\mathbb{X}_R^{I_1} ... \bar\de^{\ell_n} \mathbb{X}_R^{I_n}$ with
$N_R = \sum_i \ell_i$ in exponential form \be \bar\de^{\ell_1}
\mathbb{X}_R^{I_1} ... \bar\de^{\ell_n} \mathbb{X}_R^{I_n} =
\left[{\de\over \de \beta^{(\ell_1)}_{I_1}} ... {\de\over
\de\beta^{(\ell_n)}_{I_n}} \exp{\sum_k \beta^{(\ell)}_I
\bar\de^\ell \mathbb{X}_R^I}\right]_{\beta^{(\ell)}_I =0} \ee one
gets

\begin{eqnarray}
\cW_R(\bar{z}_i) = \bar{z}_{14}^{- {{}} |{\bf p}_R|^2} \cI_R(p_i,
\bar{z}_i) \left[\ \Phi^{(1)}_{...}
\left(\frac{\partial}{\partial\beta^{(1)}}...\right)_{N_R}{\de\over
\de b_2^M} {\de\over \de b_3^N} \Phi^{(4)}_{...}
\left(\frac{\partial}{\partial\beta^{(4)}}...\right)_{N_R} \right.
\nn \\ \left. \cU_R(\bar{z}_i)
\exp\left({\sum_{n,m}(-)^n(n+m-1)!{\beta^{(1)}_n\cdot\beta^{(4)}_n
\over \bar{z}_{14}^{n+m}}}\right)
\right]_{^{b_2=b_3=0}_{\beta^{(1)}_n= \beta^{(4)}_m=0}}
\end{eqnarray}
where, similarly to $\cI_L(p_i, {z}_i)$,  \be \cI_R(p_i,
\bar{z}_i) = \prod_{i<j} \bar{z}_{ij}^{ {{}}p_i p_j} \ee is the
R-mover Koba-Nielsen factor and
\begin{eqnarray}
\cU_R(\bar{z}_i)&=&\frac{(b_2b_3)}{z_{23}^2}+\left(\sum_n\frac{(-)^n
n!b_2\cdot\beta^{(1)}_n}{z_{12}^{n+1}}
+\sum_m\frac{m!b_2\cdot\beta^{(4)}_m}{z_{24}^{m+1}}+
i \sum_{i\neq2}\frac{b_2\cdot \mathbb{P}_{R,i}}{z_{2i}}\right) \nonumber \\
& & \times \left(\sum_n\frac{(-)^n
n!b_3\cdot\beta^{(1)}_n}{z_{13}^{n+1}}+
\sum_m\frac{m!b_3\cdot\beta^{(4)}_m}{z_{34}^{m+1}}+ i
\sum_{j\neq3}\frac{b_3\cdot\mathbb{P}_{R,j}}{z_{3j}}\right)
\end{eqnarray}
Factoring out $\bar{z}_{23}^{-2}\bar{z}_{14}^{-2N_R}$, one
eventually finds \be \cW_R(\bar{z}_i) =
\langle\Phi_1|\Phi_4\rangle({\bf p}_R) \cY_R({\bf p}_R, b_2,b_3;
\bar{z}) \bar{z}_{23}^{-2}\bar{z}_{14}^{-2(N_R-1) - {{}} |{\bf
p}_R|^2} \prod_{i<j} \bar{z}_{ij}^{{{}} p_i p_j} \ee that nicely
fit with the $\bar{z}_{14}^{- {{}}|{\bf p}_L|^2}\cI_L(z_i, p_i)$
L-moving factor, since $|{\bf p}_L|^2 = 2(N_R -1) + {{}} |{\bf
p}_R|^2$. The function $\cY_R({\bf p}_R, b_2,b_3; \bar{z})$ of the
cross ratio $\bar{z}=\bar{z}_{12}\bar{z}_{34}/\bar{z}_{13}
\bar{z}_{24}$ depends on the choice of colliding particles, while
$\langle\Phi_1|\Phi_4\rangle({\bf p}_R)$ denote the `overlap' of
the pair of small BH states.

\subsection{Integration and amplitudes}

Including the bosonic ghost correlator
\begin{align}
\left|\langle c(z_1)c(z_2)c(z_4)\rangle\right|^2=|z_{12}|^2 |z_{14}|^2
|z_{24}|^2
\end{align}
and setting $z_1\rightarrow \infty$, $z_2\rightarrow 1$, $z_3 = z$
and $z_4 \rightarrow 0$, one eventually gets amplitudes of the
form
\begin{eqnarray}
\cA(p_i)=g_{_{YM}}^2(2\pi)^{4}\delta(\Sigma_ip_i){{{}}}\delta^{ij}
\int d^2z \ |z|^{{{2}} k_3\cdot p_4}|1-z|^{{{2}}k_2\cdot k_3-4}
\mathcal{E}_L(z)\ \mathcal{E}_R(\bar{z})
\end{eqnarray}
where
\begin{eqnarray}
\mathcal{E}_L=(a_2a_3)({{}}k_2k_3-1)-{{}}(a_2k_3)(p_4a_3)\frac{1-z}{z}
+{{}}(a_2p_4)(k_2a_3)(1-z)-{{}}(a_2p_4)(p_4a_3)\frac{(1-z)^2}{z} \nonumber\\
\end{eqnarray}
and $\mathcal{E}_R$ depends on the incoming particles. All the
necessary integrals are of the form \cite{Gross:1985fr,
Gross:1985rr}
\begin{eqnarray}\label{I(a,n,b,m)} I(a,n,b,m)&=&\int
d^2z|z|^a|1-z|^bz^n(1-z)^m \nonumber \\
&=&\frac{\Gamma(-1-(a+b)/2)\Gamma(1+n+a/2)\Gamma(1+m+b/2)}
{\Gamma(-a/2)\Gamma(-b/2)\Gamma(2+n+m+(a+b)/2)}
\end{eqnarray}
and lead to the Shapiro-Virasoro-like form factor \be
\mathcal{F}_{SV} =
\frac{\Gamma(1+{{}}k_2k_3)\Gamma({{}}k_2p_1)\Gamma({{}}k_2p_4)}
{\Gamma(2-{{}}k_2k_3)\Gamma(-{{}}k_2p_1)\Gamma(-{{}}k_2p_4)}
\label{SVformfact}\ee up to rational expressions in the
kinematical variables.

In the following, for illustrative purposes, we will specialize to
the case $N_R =2$. Generalization to higher level is tedious but
straightforward using the above procedure for the scalar product
of `internal' R-moving states.

\subsection{2 Vectors - 2 small BH amplitude: mutually neutral case}

We first consider the case in which the two charged small BH's are
neutral wrt to the incoming gauge bosons \ie they are charged wrt
a `hidden' gauge group not the `visible' one. The tree-level
amplitude for this process reads
\begin{equation}
\cA^{ij,a,b,kl}_{vv\rightarrow \Phi\bar\Phi}(p_i)=g_s^2\int
d^2z_3\langle c\bar{c}V_{\Phi}^{ij(-1)}(z_1,\bar{z}_1)
c\bar{c}V^{a(0)}_{v}(z_2,\bar{z}_2) V^{b(0)}_{v}(z_3,\bar{z}_3)
c\bar{c}V_{\bar{\Phi} }^{kl(-1)}(z_4,\bar{z}_4) \rangle
\end{equation}
where
\begin{eqnarray}
\label{VphiNoch}
V_{\Phi}^{ij(-1)}&=&\psi^i e^{-\varphi}e^{i{\bf p}_L{\bf X}_L}\bar{\partial}^2X_R^j
e^{i{\bf p}_R{\bf X}_R}e^{ip\cdot X}
\end{eqnarray}
describes a small BPS BH with mass $M^2=|{\bf p}_L|^2=|{\bf
p}_R|^2+{{2}}$ ($N_R=2$). $\cA^{ij,a,b,kl}_{vv\rightarrow
\Phi\bar\Phi}$ can be decomposed as
\begin{eqnarray}
&&\cA^{ij,a,b,kl}_{vv\rightarrow \Phi\bar{\Phi}}(p_i)=g_s^2\int
d^2z_3|z_{12}|^2|z_{14}|^2|z_{24}|^2 \langle
e^{-\varphi(z_1)}e^{-\varphi(z_4)}\rangle \langle
\psi^i(z_1)\psi^k(z_4)\rangle
\nonumber \\
& &\langle e^{i{\bf p}_L{\bf X}_L(z_1)}e^{-i{\bf p}_L {\bf
X}_L(z_4)}\rangle\langle\bar{\partial}^2X_R^j e^{i{\bf p}_R {\bf
X}_R}(\bar{z}_1)\bar{\partial}^2X_R^l e^{-i{\bf p}_R {\bf
X}_R}(\bar{z}_4)\rangle \langle
\bar{J}^a(\bar{z}_2)\bar{J}^b(\bar{z}_3)\rangle
\nonumber \\
&&\langle e^{ip\cdot X}(z_1)a_{\mu}\left(\partial X^\mu+ik \cdot \psi\psi^\mu\right)e^{ik\cdot X}(z_2)
a_{\rho}\left(\partial X^\rho+ik \cdot
\psi\psi^\rho\right)e^{ik\cdot X}(z_3)e^{ip\cdot
X}(z_4)\rangle
\end{eqnarray}

The R-moving contribution $\cE_R$ consists in the internal boson
correlator \be \langle\bar{\partial}^2X_R^j e^{i{\bf p}_R{\bf
X}_R}(\bar{z}_1) \bar{\partial}^2X_R^l e^{-i{\bf p}_R {\bf
X}_R}(\bar{z}_4) \rangle =
{{}}\left(6\frac{\delta^{jl}}{\bar{z}_{14}^4}-{{}}\frac{p_R^jp_R^l}{\bar{z}_{14}^4}\right)
\bar{z}_{14}^{-{{}}|{\bf p}_R|^2} \ee and in the current
correlator
\begin{align}
\langle
\bar{J}^a(\bar{z}_2)\bar{J}^b(\bar{z}_3)\rangle=\frac{\delta^{ab}}{\bar{z}_{23}^2}
\end{align}

Reinstating normalization factors, one finally gets
\begin{align} \cA^{ij,a,b,kl}_{vv\rightarrow \Phi\bar\Phi}(p_i) &
={g_{_{YM}}^2\over M_s^2}(2\pi)^{4}\delta(\Sigma_ip_i){{}}
(\tilde{a}_2\tilde{a}_3)
\mathcal{F}_{SV}F\delta^{ik}\delta^{ab}\left(6\delta^{jl}-{{}}p_R^jp_R^l\right)
\end{align}
where the kinematical factor $F$ and the Shapiro-Virasoro-like
form factor $\mathcal{F}_{SV}$ are defined in (\ref{KinFactF}) and
(\ref{SVformfact}) while $\tilde{a}_{i\mu}$ are manifestly gauge
invariant polarizations defined in Eq (\ref{aTilde}).

It is worth noticing that in the `formal' field theory limit
$\ap\rightarrow 0$ ($M_s \rightarrow \infty$) with fixed $M$,
$\mathcal{F}_{SV}\rightarrow 1$ and
$\cA^{ij,a,b,kl}_{vv\rightarrow \Phi\bar\Phi}(p_i)$ reproduces the
supergravity result. To lowest order the process is indeed
mediated by graviton exchange that is suppressed by a factor of
$g_{_{YM}}^2/M_s^2 \sim 1/M_{Pl}^2$.

For heterotic strings, one can replace $\de^2 X^k$ in one or both
vertex operators with $\de X^i \de X^j$. These states mix with
each other and contribute to the degeneracy of (very) small BH's
with $N_R=2$.

\subsection{2 Vectors - 2 small BH amplitude: mutually charged case}

We now consider the more interesting but slightly more involved
case in which the two small BH's are charged wrt the `visible'
gauge group of the two incoming vector bosons. The tree-level
amplitude for this process reads
\begin{equation}
\cA^{ika,b,c,jld}_{vv\rightarrow \Phi\bar\Phi}(p_i)=g_s^2\int
d^2z_3\langle c\bar{c}V_{\Phi}^{ika(-1)}(z_1,\bar{z}_1)
c\bar{c}V_v^{b(0)}(z_2,\bar{z}_2) V_v^{c(0)}(z_3,\bar{z}_3)
c\bar{c}V_{\bar\Phi}^{jld(-1)} (z_4,\bar{z}_4)\rangle
\end{equation}
where
\begin{eqnarray}
V_{\Phi}^{ika(-1)} &=&\psi^i e^{-\varphi}e^{i{\bf p}_L{\bf
X}_L}\bar{\partial}X_R^k \bar{J}^a
e^{i{\bf p}_R{\bf X}_R}e^{ip\cdot X}
\end{eqnarray}
In the R-moving sector, in addition to the elementary two-point
function \be \langle \bar{\partial}X_R^k \bar{\partial}X_R^l
\rangle = - {{}} {\delta^{kl} \over \bar{z}_{14}^2} \ee one needs
the correlator of four currents given by
\begin{eqnarray}
\langle
\bar{J}^{a_1}(\bar{z}_1)\bar{J}^{a_2}(\bar{z}_2)\bar{J}^{a_3}(\bar{z}_3)\bar{J}^{a_4}(\bar{z}_4)\rangle
&=& \left[\frac{T_{12}T_{34}}{\bar{z}_{12}^2\bar{z}_{34}^2}+\frac{T_{13}T_{24}}{\bar{z}_{13}^2\bar{z}_{24}^2}+\frac{T_{14}T_{23}}{\bar{z}_{14}^2\bar{z}_{23}^2}\right]\nonumber \\
&+&2\left[\frac{T_{[12][34]}}{\bar{z}_{12}\bar{z}_{23}\bar{z}_{34}\bar{z}_{41}}+\frac{T_{[13][24]}}{\bar{z}_{13}\bar{z}_{32}\bar{z}_{24}\bar{z}_{41}}\right]
\end{eqnarray}
where $T_{ij} = \Tr(T_{a_i}T_{a_j})$ and  $T_{[ij][kl]} = \Tr([T_{a_i},T_{a_j}][T_{a_k},T_{a_l}])$.\\
This leads to
\begin{eqnarray}
\mathcal{E}_R(\bar{z})&=&T_{14}T_{23}+T_{12}T_{34}\left(1-\frac{2}{\bar{z}}+\frac{1}{\bar{z}^2}\right)+T_{13}T_{24}\left(1-2\bar{z}+\bar{z}^2\right) \nonumber \\
&& +{}2T_{[12][34]}\left(1 - {1\over \bar{z}}\right)+
2T_{[13][42]}(1-\bar{z})
\end{eqnarray}
that combined with $\cE_L(z)$ and integrated yields
\begin{eqnarray}
\cA^{ika,b,c,jld}_{vv\rightarrow \Phi\bar\Phi}(p_i) =
g_s^2(2\pi)^{4}\delta(\Sigma_ip_i)
{{{}}}\delta^{ij}\left(-\delta^{kl}-{{}}p_R^kp_R^l\right)
\mathcal{J}
\end{eqnarray}
where
\begin{eqnarray}
\mathcal{J}&=&A_0\left(T_{12}T_{34}+T_{13}T_{24}+T_{14}T_{23}+
2T_{[13][24]}+ 2T_{[12][34]}\right)+A_2T_{13}T_{24}
\\ \nonumber
&-&2A_1\left(T_{13}T_{24}+
T_{[13][24]}\right)-2A_{-1}\left(T_{12}T_{34} +
T_{[12][34]}\right)+A_{-2}T_{12}T_{34}
\end{eqnarray}
with
\begin{eqnarray}
\label{An}
A_n=\int d^2 z \ |z|^{2k_3p_4}|1-z|^{2k_2k_3-4} \ \bar{z}^n \
\mathcal{E}_L(z) = (\tilde{a_2}\tilde{a_3}) \ F \
\frac{\Gamma(1+k_2k_3)}{\Gamma(2-k_2k_3)} \ J_n
\end{eqnarray}
where
\begin{eqnarray}
&&J_{0}=-\frac{\Gamma(k_3p_4)\Gamma(k_3p_1)}{\Gamma(-k_3p_4)\Gamma(-k_3p_1)}
\\
&&J_{1}=-\frac{1+k_3p_4}{k_3p_1}J_0, \ \ \ \
J_{-1}=-\frac{1+k_3p_1}{k_3p_4}J_0 \\
&&J_{2}=\frac{(2+k_3p_4)(1+k_3p_4)}{(k_3p_1-1)k_3p_1}J_0, \ \ \ \
J_{-2}=\frac{(2+k_3p_1)(1+k_3p_1)}{(k_3p_4-1)k_3p_4}J_0
\end{eqnarray}
Reinstating normalization factors, one eventually gets
\begin{eqnarray}
\label{Acharged} \cA^{ika,b,c,jld}_{vv\rightarrow
\Phi\bar\Phi}(p_i) = {g_{_{YM}}^2\over M_s^2}
(2\pi)^{4}\delta(\Sigma_ip_i)
{{{}}}\delta^{ij}\left(\delta^{kl}+{{}}p_R^kp_R^l\right)(\tilde{a}_2\tilde{a}_3)F\cal{F}_{SV}\mathcal{I}
\end{eqnarray}
where
\begin{eqnarray}
\mathcal{I}&=&T_{12}T_{34}+T_{13}T_{24}+T_{14}T_{23}+2T_{[13][24]}+2T_{[12][34]} \nonumber \\
&&+\frac{2(2+\ap k_3p_4)}{\ap k_3p_1}[T_{13}T_{24}+
T_{[13][24]}]+\frac{2(2+\ap k_3p_1)}{\ap k_3p_4}\left[T_{12}T_{34}+T_{[12][34]}\right] \nonumber \\
&& +\frac{(4+\ap k_3p_1)(2+\ap k_3p_1)}{\ap k_3p_4(2-\ap
k_3p_4)}T_{12}T_{34} +\frac{(4+\ap k_3p_4)(2+\ap k_3p_4)}{\ap
k_3p_1(2-\ap k_3p_1)}T_{13}T_{24}
\end{eqnarray}

In the `formal' field theory limit $\ap\rightarrow 0$
($M_s\rightarrow \infty$) with fixed $M$,
$\mathcal{F}_{SV}\rightarrow 1$, only terms with `poles' in $\cI$
survive and the amplitude reproduces the SYM theory result.
Similarly to the previous case (BPS BH's charged wrt to a `hidden'
sector), one can replace $\de X^k J^a$ with $\de J^a$ in one or
both BPS vertex operators. This would lead to mixing and
eventually account for the degeneracy of the small BH's.

\subsection{2 Gravitons - 2 small BH's amplitude}

Finally we consider small BH pair production in high energy
graviton collisions. This process is practically impossible at LHC
or near future accelerators but it may prove dominant in the very
early universe or in models with low-scale gravity. The tree-level
amplitude for the process reads
\begin{equation}
\cM^{ij,kl}_{hh\rightarrow \Phi\bar\Phi}(p_i) =g_s^2\int
d^2z_3\langle c\bar{c}V_{\Phi}^{ij(-1)}(z_1,\bar{z}_1)
c\bar{c}V_h^{(0)}(z_2,\bar{z}_2) V_h^{(0)}(z_3,\bar{z}_3)
c\bar{c}V_{\bar\Phi}^{kl(-1)} (z_4,\bar{z}_4)\rangle
\end{equation}
where $V_{\Phi}^{ij(-1)}$ has been defined in Eq.
(\ref{VphiNoch}). For calculation purposes, it is convenient to
`factorize' graviton polarization tensors as
$h_{\mu\nu}^{(2\lambda)}= a_\mu^{(\lambda)} a_\nu^{(\lambda)}$,
that satisfy $k^\mu h_{\mu\nu}=h_{\mu\nu}k^\nu=0$ and
$h^\mu_{\phantom{\mu}\mu}=0$. The amplitude can be decomposed as
\begin{eqnarray}
&&\cM^{ij,kl}_{hh\rightarrow \Phi\bar\Phi}(p_i) =g_s^2\int
d^2z_3|z_{12}|^2|z_{14}|^2|z_{24}|^2\rangle \langle
e^{-\varphi(z_1)}e^{-\varphi(z_4)}\rangle
\langle \psi^i(z_1)\psi^k(z_4)\rangle \nonumber \\
&& \langle e^{i{\bf p}_L{\bf X}_L}(z_1) e^{-i{\bf p}_L{\bf
X}_L}(z_4)\rangle
\langle \bar{\partial}^2X_R^j e^{i{\bf p}_R{\bf X}_R}(z_1)\bar{\partial}^2X_R^l e^{-i{\bf p}_R{\bf X}_R}(z_4)\rangle \nonumber \\
&&\langle e^{ip\cdot X}(z_1) a_2^\nu \bar{\partial}X_\nu(z_2)
a_3^\sigma \bar{\partial}X_\sigma(z_3) e^{ip\cdot X}(z_4)\rangle
\nonumber \\
&& \langle e^{ip\cdot X}(z_1)a_2^{\mu}\left(\partial X_\mu+ik
\cdot \psi\psi_\mu\right) e^{ik\cdot X}(z_2)
a_3^{\rho}\left(\partial X_\rho+ik \cdot
\psi\psi_\rho\right)e^{ik\cdot X}(z_3) e^{ip\cdot X} (z_4)\rangle
\end{eqnarray}

The R-moving contribution requires \be {\cB}^{s-t}_R = \langle
e^{ip_1\cdot X}(\bar{z}_1) a_2^\nu \bar{\partial}X_\nu
e^{ik_2\cdot X}(\bar{z}_2) a_3^\sigma \bar{\partial}X_\sigma
e^{ik_3\cdot X}(\bar{z}_3)e^{ip_4\cdot X}(\bar{z}_4) \rangle \ee
Eventually the result is simply given by ${\cB}^{s-t}_R (\bar{z})
=\cB^{s-t}_{L}(z)_{z\rightarrow \bar{z}}$ previously computed in
Eq.(\ref{Bs-t}).

Combining L- and R-moving parts one has
\begin{align} \cM^{ij,kl}_{hh\rightarrow \Phi\bar\Phi}(p_i) &
=g_s^2(2\pi)^{4}\delta(\sum_ip_i){{{}}}
\delta^{ik}\left(6\delta^{jl}-{{}}p_R^jp_R^l\right)\mathcal{W}
\end{align}
where $\mathcal{W}$ can be expressed in terms of the integrals
$I(a,n,b,m)$ in (\ref{I(a,n,b,m)}). Setting $I(a,0,b,0)=I_{0}$,
using the factorial properties of $\Gamma$ function and the
compact notation \be hh=h_{\mu\nu}h^{\mu\nu}\quad, \quad php=p^\mu
h_{\mu\nu}p^\nu\quad, \quad phhp=p^\mu
h_{\mu\nu}h^{\nu\sigma}p_\sigma\ee one finds

\begin{align}
\label{W} \frac{\mathcal{W}}{I_0({{1}}-k_2k_3)} &
=h_2h_3{{}}-k_3h_2h_3k_2-\frac{\left[(2{{}}-k_2k_3)(k_3h_2h_3p_4)-(k_3h_2k_3)(p_4h_3k_2)\right]}{(k_3p_4)} \nonumber \\
&
-\frac{\left[(2{{}}-k_2k_3)(p_4h_2h_3k_2)-(k_3h_2p_4)(k_2h_3k_2)\right]}{(k_2p_4)} \nonumber \\
&
+\frac{(k_2k_3)\left[(2{{}}-k_2k_3)(p_4h_2h_3p_4)-(p_4h_2k_3)(k_2h_3p_4)\right]}{(k_3p_4)(k_3p_1)} \nonumber \\
&
+\frac{({{1}}-k_2k_3)(k_3h_2k_3)(p_4h_3p_4)}{(k_3p_4)^2}+\frac{2({{1}}-k_2k_3)(k_3h_2p_4)(p_4h_3k_2)}{(k_3p_4)(k_2p_4)} \nonumber \\
&
+\frac{({{1}}-k_2k_3)(k_2h_3k_2)(p_4h_2p_4)}{(k_2p_4)^2}-\frac{2({{1}}-k_2k_3)(k_2k_3)(k_2h_3p_4)(p_4h_2p_4)}{(k_2p_4)^2(k_3p_4)} \nonumber \\
&
+\frac{({{1}}-k_2k_3)(k_2k_3)^2(p_4h_2p_4)(p_4h_3p_4)}{(k_3p_4)^2(k_2p_4)^2}-\frac{2({{1}}-k_2k_3)(k_2k_3)(p_4h_3p_4)(k_3h_2p_4)}{(k_3p_4)^2(k_2p_4)}
\end{align}

The integral $I_0$ is given by \be I_0 = {
\mathcal{F}_{SV}{F}\over k_2k_3 -{{1}}} \ee where $F$ and
$\mathcal{F}_{SV}$ are the by now familiar kinematical factor and
S-V form factor. One eventually gets
\begin{align}
\cM^{ij,kl}_{hh\rightarrow \Phi\bar\Phi}(p_i) =\frac{16\pi}{
M_{Pl}^2} \
(2\pi)^{4}\delta(\Sigma_ip_i){{{}}}\mathcal{F}_{SV}F\delta^{ik}
\left(6\delta^{jl}-{{}}p_R^jp_R^l\right)\left[(\tilde{h}_2\tilde{h}_3)+{{}}\cH
\right]
\end{align}
where
\begin{align}
\tilde{h}_{i\mu\nu}=\left(\delta_\mu^{\phantom{\mu}\rho}-\frac{k_{i\mu}
p_4^\rho}{p_4k_i}\right)\left(\delta_\nu^{\phantom{\nu}\sigma}-\frac{k_{i\nu}
p_4^\sigma}{p_4k_i}\right)h_{\rho\sigma}
\end{align}
is a manifestly gauge invariant quantity and
\begin{eqnarray}
&&\cH ={\ap \over 2}
\left\{-(k_2k_3)(\tilde{h}_2\tilde{h}_3)+(k_2k_3)(h_2h_3)-(k_3h_2h_3k_2) \right.\\
&& \left. + \frac{(k_2k_3)(p_1h_2h_3p_4)-
(p_4h_3k_2)(p_1h_2k_3)}{k_3p_4}+
\frac{(k_2k_3)(p_1h_2h_3p_4)-(p_4h_2k_3)(p_1h_3k_2)}{k_2p_4}\right\}
\nn
\end{eqnarray} represent higher-derivative $\alpha'$ corrections.
In the `formal' field theory limit $\ap\rightarrow 0$,
$\mathcal{F}_{SV}\rightarrow 1$, the surviving gravitational
amplitude $\tilde{h}_2\tilde{h}_3$ is essentially the square of
the gauge theory amplitude $\tilde{a}_2\tilde{a}_3$.

As in the first case considered, one can replace $\de^2 X^k$ in
one or both BPS scalar vertex operators with $\de X^i \de X^j$.
These states mix with each other and contribute to the degeneracy
of the resulting `small BH'.

\section{Cross section, Angular and Energy distribution}
\label{crossect}

 We are now ready to compute the cross section for
pair production of small BH's in graviton or gauge boson
scattering.

With respect to field theory amplitudes for pair production of
massive (charged) scalars, heterotic string amplitudes contain
higher derivative correction and are dressed with
Shapiro-Virasoro-like form factors defined in (\ref{SVformfact}),
that contain further higher-derivative corrections. Moreover,
summming over final BPS states with the same charges and mass but
different (unresolved) R-moving string oscillator modes enhances
the result by the micro-state degeneracy factor $d(N_R)$. For
large $N_R$, $\log d(N_R)\approx 4\pi\sqrt{N_R}$ reproduces the
`macro-scopic' Wald entropy of the small BH's.

It is easy to check that gravity mediated amplitudes, including
the one with product BH's neutral wrt the incoming vector bosons,
are largely suppressed wrt the amplitudes with products BH's
charged wrt to the incoming gauge bosons. In more realistic
scenari small BH's that couple minimally to the `visible' gauge
group have a chance to be produced even at LHC
\cite{Gingrich:2009hj}.

To proceed further, let us recall that in the CM frame Mandelstam
variables assume values \begin{eqnarray}
s&=&-(k_2+k_3)^2=-2(k_2k_3)=4E^2 \\
t&=&-(k_2+p_1)^2=M^2-2(k_2p_1)=M^2-2E^2(1+\sqrt{1-\mu^2}\cos\theta) \\
u&=&-(k_2+p_4)^2=M^2-2(k_2p_4)=M^2-2E^2(1-\sqrt{1-\mu^2}\cos\theta)
\end{eqnarray}
where $\mu=M/E = 1/\eta$. Exploiting the notation
$\hat{w}=\alpha'w/4$, the S-V form factor (\ref{SVformfact}) reads
\begin{eqnarray}
\mathcal{F}_{SV}=\frac{\Gamma(1-\hat{s})\Gamma(\hat{M}^2-\hat{t})\Gamma(\hat{M}^2-\hat{u})}
{\Gamma(2+\hat{s})\Gamma(\hat{t}-\hat{M}^2)\Gamma(\hat{u}-\hat{M}^2)}
\end{eqnarray}
By using the factorial property of $\Gamma$ function and
$\Gamma(z)\Gamma(1-z)=\pi/\sin(\pi z)$, one gets
\begin{eqnarray}
\mathcal{F}_{SV}=\frac{\hat{s}}{(1+\hat{s})}\frac{\sin(\pi
\hat{s})}{\pi} \
B(-\hat{s},\hat{M}^2-\hat{t})B(-\hat{s},\hat{M}^2-\hat{u})
\end{eqnarray}
Then, using the Mittag-Leffler expansion of $B(u,v)$
\begin{eqnarray}
B(u,v)=\sum_{n=0}^\infty \frac{\cR_n(u)}{v+n}
\end{eqnarray}
where $\cR_n(u)=(-1)^n(u-1)\ldots (u-n)/n!$, one obtains
\begin{eqnarray}
\mathcal{F}_{SV}=\frac{\hat{s}}{(1+\hat{s})}\frac{\sin(\pi
\hat{s})}{\pi}\sum_{n=0}^\infty\frac{\cR_n(\hat{s})}{(a_n+bx)}\sum_{k=0}^\infty\frac{\cR_k(\hat{s})}{(a_k-bx)}
\end{eqnarray}
where $a_n=n+\hat{s}/2$, $b=(\hat{s}/2)\sqrt{1-\mu^2}$ and $x=\cos\theta$. \\

\subsection{Cross section for small BH's in the `visible' sector}

Henceforth we will focus on pair production of small BH's charged
wrt to the `visible' gauge group, whose transition amplitude is
given by (\ref{Acharged}) {\it viz.}
\begin{eqnarray}
|\cA|^2 = {g_{_{YM}}^4 \over M_s^4}
|\mathcal{F}_{SV}|^2F^2\mathcal{I}^2(\tilde{a}_2\tilde{a}_3)^2
\end{eqnarray}
where
\begin{eqnarray}
\mathcal{I}&=&T_{12}T_{34}+T_{13}T_{24}+T_{14}T_{23}+2T_{[13][24]}+2T_{[12][34]} \nonumber \\
&&+\frac{2(a_1+bx)}{(a_0-bx)}(T_{13}T_{24}+ T_{[13][24]})+
\frac{2(a_1-bx)}{(a_0+bx)}(T_{12}T_{34}+T_{[12][34]})\nn \\
&&+\frac{(a_2-bx)(a_1-bx)}{(a_{-1}+bx)(a_0+bx)}T_{12}T_{34}+
\frac{(a_2+bx)(a_1+bx)}{(a_{-1}-bx)(a_0-bx)}T_{13}T_{24}
\end{eqnarray}
and \be F= {1\over 2}E^2 [(1-\mu^2) x^2 - 1] \ee
 In the helicity basis, the amplitude reads
\begin{eqnarray}
|\cA|^2 = {g_{_{YM}}^4 \over M_s^4}
|\mathcal{F}_{SV}|^2F^2\mathcal{I}^2[(1+f)^2\sigma_{\lambda_1,
\lambda_2}+f^2\delta_{\lambda_1, \lambda_2}]
\end{eqnarray}
Averaging over helicities of the incoming vector bosons and
summing over final states\footnote{Including superpartners of the
scalar states would be tantamount to replacing a factor of $6$
with a factor of $16 = 8_B + 8_F$.} of the small scalar BH's one
gets
\begin{eqnarray}
\label{Apoli} &&\langle |\cA|^2\rangle = {3 g_{_{YM}}^4 \over
M_s^4}\ d(N_R)^2\
\mathcal{F}_{SV}^2F^2 \mathcal{I}^2(1+2f+2f^2) \nn \\
&&\quad ={3 g_{_{YM}}^4 \over 4 M_s^4}\ d(N_R)^2 \frac{\hat{s}^2 \
\sin^2(\pi\hat{s})}{\pi^2(1+\hat{s})^2}\left|\sum_{n=0}^\infty\frac{\cR_n(\hat{s})}
{(a_n+bx)}\right|^2\left|\sum_{k=0}^\infty\frac{\cR_k(\hat{s})}
{(a_k-bx)}\right|^2\mathcal{I}^2\left[D(1-x^2)^2+H\right] \nn\\
\end{eqnarray}
where $D=(E^2-M^2)^2$ and $H=M^4$.\\
Setting \be \cT_1 =
T_{12}T_{34}+T_{13}T_{24}+T_{14}T_{23}+2T_{[13][24]}+2T_{[12][34]}
\quad , \ee \be \cT_2 = 2(T_{13}T_{24}+ T_{[13][24]})\quad , \quad
\cT_3=2(T_{12}T_{34}+T_{[12][34]})\quad , \ee \be \cT_4 =
T_{12}T_{34} \quad , \quad \cT_5 = T_{13}T_{24} \ee one finds
\begin{eqnarray} &&
\cI^2=\mathcal{T}_1^2+\frac{(a_1+bx)^2}{(a_0-bx)^2}\mathcal{T}_2^2+\frac{(a_1-bx)^2}{(a_0+bx)^2}\mathcal{T}_3^2+\frac{(a_2-bx)^2(a_1-bx)^2}{(a_{-1}+bx)^2(a_0+bx)^2}\mathcal{T}_4^2\nonumber \\
&&+\frac{(a_2+bx)^2(a_1+bx)^2}{(a_{-1}-bx)^2(a_0-bx)^2}\mathcal{T}_5^2+2\mathcal{T}_1
\mathcal{T}_2\frac{(a_1+bx)}{(a_0-bx)}+2\mathcal{T}_1
\mathcal{T}_3\frac{(a_1-bx)}{(a_0+bx)}\nonumber \\
&&+2\mathcal{T}_1
\mathcal{T}_4\frac{(a_2-bx)(a_1-bx)}{(a_0+bx)(a_{-1}+bx)}+2\mathcal{T}_1
\mathcal{T}_5\frac{(a_2+bx)(a_1+bx)}{(a_0-bx)(a_{-1}-bx)}\nonumber \\
&&+2\mathcal{T}_2
\mathcal{T}_3\frac{(a_1+bx)(a_1-bx)}{(a_0+bx)(a_0-bx)}+2\mathcal{T}_2
\mathcal{T}_4\frac{(a_1+bx)(a_1-bx)(a_2-bx)}{(a_0+bx)(a_0-bx)(a_{-1}+bx)}\nonumber \\
&&+2\mathcal{T}_2
\mathcal{T}_5\frac{(a_1+bx)^2(a_2+bx)}{(a_0-bx)^2(a_{-1}-bx)}+2\mathcal{T}_3
\mathcal{T}_4\frac{(a_1-bx)^2(a_2-bx)}{(a_0+bx)^2(a_{-1}+bx)}\nonumber \\
&&+2\mathcal{T}_3
\mathcal{T}_5\frac{(a_1+bx)(a_1-bx)(a_2+bx)}{(a_0+bx)(a_0-bx)(a_{-1}-bx)}+2\mathcal{T}_4
\mathcal{T}_5\frac{(a_1+bx)(a_1-bx)(a_2+bx)(a_2-bx)}{(a_0+bx)(a_0-bx)(a_{-1}+bx)(a_{-1}-bx)}\nonumber \\
\end{eqnarray}
In order to average over colours one can peruse the relation
$Tr_R(T^aT^b)=\ell_R\delta^{ab}$ and its corollary $\ell_R d_G =
C_R d_R$, where $d_G$ is the dimension of the group, $d_R$ the
dimension of the representation $R$ and $C_R$ its (quadratic)
Casimir. For the Adjoint representations of both $E(8)$ and
$SO(32)$ $C_A =30$. Setting $\langle \cT\rangle_c=\cT/d^2_G$, one
eventually gets
\begin{eqnarray}
&&\langle\mathcal{T}_1^2\rangle_c=\frac{12C^2_A}{d_G}+16C_Ad_G+\frac{6}{d_G}+3,
\quad\langle\mathcal{T}_2^2\rangle_c=\langle\mathcal{T}_3^2\rangle_c=\frac{4C^2_A}{d_G}+4 \\
&&\langle\mathcal{T}_4^2\rangle_c=\langle\mathcal{T}_5^2\rangle_c=1,
\quad\langle\mathcal{T}_1\mathcal{T}_2\rangle_c=\langle\mathcal{T}_1\mathcal{T}_3\rangle_c
=\frac{6C^2_A}{d_G}+8C_Ad_G+\frac{4}{d_G}+2 \\
&&\langle\mathcal{T}_1\mathcal{T}_4\rangle_c=\langle\mathcal{T}_1\mathcal{T}_5\rangle_c
=2C_Ad_G+\frac{2}{d_G}+1,
\quad\langle\mathcal{T}_2\mathcal{T}_3\rangle_c=\frac{2C^2_A}{d_G}+8C_Ad_G+\frac{4}{d_G}\\
&&\langle\mathcal{T}_2\mathcal{T}_4\rangle_c=\langle\mathcal{T}_3\mathcal{T}_5\rangle_c
=2C_Ad_G+\frac{2}{d_G}, \quad
\langle\mathcal{T}_2\mathcal{T}_5\rangle_c=\langle\mathcal{T}_3\mathcal{T}_4\rangle_c=2,
\quad\langle\mathcal{T}_4\mathcal{T}_5\rangle_c=\frac{1}{d_G}
\end{eqnarray}
Reinstating normalization factors, the differential cross section
becomes
\begin{eqnarray}
\frac{d\sigma}{d\Omega}&=&\frac{3g_{_{YM}}^4 d(N_R)^2 }{(8\pi)^4
M_s^2}  \sqrt{1-\mu^2}
\left[(1-\mu^2)^2(1-x^2)^2+\mu^4\right]\langle\cI^2\rangle_c \nn \\
&&\frac{\hat{s}^3
\sin^2(\pi\hat{s})}{(1+\hat{s})^2}\left|\sum_{n=0}^\infty\frac{\cR_n(\hat{s})}
{(a_n+bx)}\right|^2\left|\sum_{k=0}^\infty\frac{\cR_k(\hat{s})}{(a_k-bx)}\right|^2
\end{eqnarray}
that displays -- albeit not very explicitly -- the angular
distribution of the products.

In order to compute the total cross section one should perform
integrals of the form (\eg for the colour structure
$\mathcal{T}_1^2$)
\begin{eqnarray}
&&\int_{-1}^{1}dx \ \frac{D(1-x^2)^2+H}{(a_n+bx)(a_{n'}+bx)(a_k-bx)(a_{k'}-bx)}\nonumber \\
&&=\int_{-1}^{1}dx\left\{D\left[\frac{C_n}{(a_n+bx)}+\frac{C_{n'}}{(a_{n'}+bx)}+\frac{C_k}{(a_k+bx)}+\frac{C_{k'}}{(a_{k'}+bx)}\right]+H\right\}\nonumber \\
&&= D \log\left[\left(\frac{a_n+b}{a_n-b}\right)^{C_n}
\left(\frac{a_{n'}+b}{a_{n'}-b}\right)^{C_{n'}}
\left(\frac{a_k+b}{a_k-b}\right)^{C_k}
\left(\frac{a_{k'}+b}{a_{k'}-b}\right)^{C_{k'}}\right]+2H
\end{eqnarray}
In a similar fashion one can compute the other integrals. We
refrain from displaying the rather uninspiring results.
Alternatively, one could derive the total cross section for pair
production by means of the optical theorem \ie computing the
forward scattering amplitude at one loop projected along BPS
states \cite{Dudas:1999gz, Chialva:2005gt,Bianchi:2006nf}. We will
not pursue this viewpoint any further here. Let us instead discuss
the energy distribution.

In addition to the obvious threshold at $E_{CM} = 2M$, the cross
section is modulated by the Regge poles, \ie string excitations.
Their presence, drastically changes the high energy behavior wrt
field theory amplitudes \cite{Kiritsis:2007zz}.

In the high energy limit ($\hat{s}>>1$), $\mu=M/E\sim 0$, one
finds \be \hat{M}^2-\hat{t} \sim {1\over 2}\hat{s}(1+x) \quad ,
\quad \hat{M}^2-\hat{u} \sim {1\over 2}\hat{s}(1-x)  \ee and \be
F^2 (1+2f+2f^2) \sim {{s}^2\over 64}(1-x^2)^2 \ee Moreover, for
fixed $n=0,\pm 1, \pm 2$, $a_n \sim b \sim \hat{s}/2$, so that
$\cI\sim \cI_\infty(x)$ becomes a rational function of
$x=\cos\theta$ independent of $s=E_{CM}^2$.

Perusing Stirling formula $\Gamma(z)\sim\sqrt{2\pi} \
z^{z-1/2}e^{-z}$ in the S-V form factor yields \be
\mathcal{F}_{SV} \sim  {\sin[\pi \hat{s} (1+x)/2] \sin[\pi \hat{s}
(1-x)/2] \over \hat{s}\sin(\pi \hat{s}/2) \cos(\pi \hat{s} /2)}
\left({2\over 1+x}\right)^{-\hat{s}(1+x)}\left({2\over
1-x}\right)^{-\hat{s}(1-x)} \ee

If one simply takes $x=0$ ($\theta=\pi/2)$, the poles at
$\hat{s}=2n$ cancel and one eventually gets
\begin{eqnarray}
\mathcal{F}_{SV} \sim \frac{2^{-2\hat{s}}}{\hat{s}} \
\tan(\pi\hat{s}/2)
\end{eqnarray}
The exponentially suppressed Regge behaviour, related to the
presence of an infinite number of string resonances, is universal
in String Theory and could mark the difference with alternative
scenari with low-scale gravity to which we now turn our attention.

\subsection{Mass scales and Large Extra Dimensions}

Following the original proposal of AADD \cite{ArkaniHamed:1998rs,
Antoniadis:1998ig}, there has been an enormous interest in models
with Large Extra Dimension and TeV scale gravity or strings. The
former predict BH production at LHC at a very high rate
\cite{Giudice:1998ck, Giddings:2001bu, Dimopoulos:2001hw,
Giudice:2001ce, Meade:2007sz, Kanti:2008eq}. The latter predict
the usual exponential decay (Regge behavior) at high energies and
the characteristic modulation by the presence of Regge poles
\cite{Dudas:1999gz, Chialva:2005gt,Bianchi:2006nf, Lust:2008qc,
Anchordoqui:2009mm}. Our results are in line with this
expectation. In particular we don't find any growth of the
scattering amplitudes with energies as in the FT toy model.
Moreover, the geometric cross section (area of BH horizon) that
sets the order of magnitude for the production of a single BH is
replaced by other dynamical quantities in pair production
processes.

In perturbative heterotic strings is notoriously difficult to
accommodate LED with coupling constants for gauge interactions.
Quite generally, tree-level coupling constants are given by
$g_s^2/\hat{V}_{int}  = g^2_{_{YM}}$ so that $M^2_s = g^2_{YM}
M^2_{Pl}$. Barring large threshold corrections, which are anyway
absent in toroidal compactifications, it seems hard if not
impossible to separate the BH mass, whose lower bound for fixed
charges is of order $M_{BH} \sim M_s \sqrt{N_R}$, from the Planck
scale so as to lower the threshold for the production process to
accessible energies. In particular, in order to have $M_{BH} \sim
M_s \sim TeV$ one should have a implausibly small gauge coupling
$g_{_{YM}} \sim 10^{-15}$.

The situation improves in theories with open and un-oriented
strings where one has instead \be g_s{\hat{V}_{\perp}\over
\hat{V}_{T^6}} = g^2_{YM} \ee with $\hat{V}_{\perp}$ the volume of
the internal space transverse to the D-branes, so that \be M^2_s =
g^4_{YM} M^2_{Pl} {\hat{V}^2_{T^6}\over \hat{V}^2_\perp} \ee which
is compatible with reasonably small $g_{_{YM}}$, low string scale
(\ie BH masses) and large extra (transverse) directions. In this
context, (small) BH's are described by bound-states of D-branes,
accommodating the `visible' sector, whose mass diverges in the
$g_s \rightarrow 0$ limit. The analysis is much more involved. So
far only static properties, such as the micro-state counting, and
the grey-body factor have been computed \cite{Maldacena:1997tm}.
Computing dynamical properties of (small) BH's corresponding to
bound-states of D-brane looks very challenging.

\section{Small BH's in the  FHSV model}
\label{BhinFHSV}

The results found for toroidal compactifications can be easily and
reliably generalized to special heterotic models with lower
supersymmetry that are still protected against significant quantum
corrections. The simplest and probably most interesting
possibility is the FHSV
 model \cite{Ferrara:1995yx}. It admits both
heterotic and Type II descriptions, related to one another by
`Second Quantized Mirror Symmetry' \cite{Ferrara:1995yx}. In the
heterotic description, the model corresponds to a freely acting
$Z_2$ orbifold of a toroidal compactification on $T^6 = T^4\times
T^2$ with identical Wilson lines on the two $E_8$ factors. The
$Z_2$ orbifold generator is \be g = \cI_4 \sigma_{\bf v} \pi_G \ee
where $\cI_4$ is the inversion of the four coordinates of $T^4$,
$\sigma_{\bf v}$ is a non-geometric order two shift along $T^2$ of
parameter $\bf v=({\bf v}_L,{\bf v}_R)$ and $\pi_G$ is the
exchange of the two $E_8$'s. Level matching requires that the
shift be left-right asymmetric with \be {\bf v}\cdot {\bf v} =
|{\bf v}_L|^2 - |{\bf v}_R|^2 = {1\over 2} ({\rm mod} \ 1) \ee

The partition function consists of four terms \be \cZ = {1\over 2}
(\cZ_{00} + \cZ_{01}+ \cZ_{10} + \cZ_{11}) \ee

In the untwisted sector one finds \be \cZ_{00} = (\cQ_o + \cQ_v)
{\Lambda_{4,4} \over |\eta^4|^2} {\Lambda^+_{2,2} \over
|\eta^2|^2} \bar{E}_8(q) \bar{E}_8(q) \ee \be \cZ_{01} = (\cQ_o -
\cQ_v) |X_o - X_v|^2 {\Lambda^-_{2,2} \over |\eta^2|^2}
\bar{E}_8(q^2)\ee where \be \bar{E}_8(q) = \sum_\a
{\theta_\a^8(0|q)\over 2 \eta^8(q)} = O_{16} + S_{16} \ee and \be
\cQ_o = V_4 O_4 - S_4 S_4 \qquad \cQ_v = O_4 V_4 - C_4 C_4 \ee
compactly represent the projection on the fermionic coordinates
\cite{Bianchi:1990tb, Bianchi:1990yu}, \be X_o + X_v = { 1 \over
\eta^4} \qquad X_o - X_v = { 4 \eta^2 \over \theta_2^2}\ee on the
bosonic coordinates of $T^4$ and \be \Lambda^+_{2,2} = \sum_{{\bf
p}_L, {\bf p}_R} q^{{{1\over 2}} |{\bf p}_L|^2} \bar{q}^{{{1\over
2}}|{\bf p}_R|^2} \qquad \Lambda^-_{2,2} = \sum_{{\bf p}_L, {\bf
p}_R} (-)^{{\bf v}_L {\bf p}_L - {\bf v}_R {\bf p}_R}  q^{{{1\over
2}}|{\bf p}_L|^2} \bar{q}^{{{1\over 2}} |{\bf p}_R|^2} \ee on the
bosonic coordinates of $T^2$.

In the twisted sector one finds \be \cZ_{10} = 16 (\cQ_s + \cQ_c)
|X_s + X_c|^2 {\tilde\Lambda^+_{2,2} \over |\eta^2|^2}
\bar{E}_8(\sqrt{q}) \ee \be \cZ_{11} = 16 (\cQ_s - \cQ_c) |X_s -
X_c|^2 {\tilde\Lambda^-_{2,2} \over |\eta^2|^2}
\bar{E}_8(-\sqrt{q})\ee where the factor $16$ accounts for the
number of fixed points and
 \cite{Bianchi:1990tb, Bianchi:1990yu} \be \cQ_s = O_4 S_4 - C_4 O_4 \qquad \cQ_c = V_4
C_4 - S_4 V_4 \ee \be X_s + X_c = { \eta^2 \over \theta_4^2}
\qquad X_s - X_c = { \eta^2 \over \theta_3^2}\ee  \be
\tilde\Lambda^+_{2,2} = \sum_{{\bf p}_L, {\bf p}_R} q^{{{1\over
2}}|{\bf p}_L+{\bf v}_L|^2} \bar{q}^{{{1\over 2}}|{\bf p}_R+{\bf
v}_R|^2} \qquad \tilde\Lambda^-_{2,2} = \sum_{{\bf p}_L, {\bf
p}_R} (-)^{{\bf v}_L {\bf p}_L - {\bf v}_R {\bf p}_R} q^{{{1\over
2}}|{\bf p}_L+{\bf v}_L|^2} \bar{q}^{{{1\over 2}} |{\bf p}_R+{\bf
v}_R|^2} \ee Due to the shift, there are no massless states in the
twisted sector.

There are however 1/2 BPS states both in the untwisted and the
twisted sector \cite{Klemm:2005pd}. Setting the left-movers in
their ground states yields\footnote{Neglecting spin carried by the
R-movers} \be \cQ_o \rightarrow {\bf 2}_v + {\bf 2}_o - {\bf 2}_s
- {\bf 2}_c \qquad {\rm vector} \ee \be \cQ_v \rightarrow {\bf
4}_o - {\bf 2}_s - {\bf 2}_c \qquad {\rm hyper} \ee \be \cQ_s
\rightarrow {\bf 2}_o - {\bf 1}_s - {\bf 1}_c \qquad {\rm half \
hyper} \ee while $\cQ_c$ only contributes excited non-BPS states.

In the untwisted sector, the surviving 1/2 BPS states are a subset
of the original BPS states in the parent $\cN = 4$ theory. \be
\cZ^{1/2 \ BPS}_{vect,untw} \quad : \quad  {1\over \bar\eta^2} \{
{\bar{X}_o} [\Gamma^+_{2,2} \bar{E}^{(2)+}_8 + \Gamma^-_{2,2}
\bar{E}^{(2)-}_8] + {\bar{X}_v} [\Gamma^+_{2,2} \bar{E}^{(2)-}_8 +
\Gamma^-_{2,2} \bar{E}^{(2)+}_8] \}\ee \be \cZ^{1/2 \ BPS}_{hyp,
untw} \quad : \quad {1\over \bar\eta^2} \{ {\bar{X}_v}
[\Gamma^+_{2,2} \bar{E}^{(2)+}_8 + \Gamma^-_{2,2}
\bar{E}^{(2)-}_8] + {\bar{X}_o} [\Gamma^+_{2,2} \bar{E}^{(2)-}_8 +
\Gamma^-_{2,2} \bar{E}^{(2)+}_8]\} \ee where \be
\Gamma^{\pm}_{2,2} = {\Lambda^+_{2,2} \pm \Lambda^-_{2,2}\over 2
\bar\eta^2} \ee and \be \bar{E}^{(2)\pm}_8 = {1\over 2}
[\bar{E}_8(q) \bar{E}_8(q) \pm \bar{E}_8(q^2)] \ee

In the twisted sector, there are new 1/2 BPS states with respect
to the original BPS states in the parent $\cN = 4$ theory. \be
\cZ^{1/2 \ BPS}_{hyp, tw}  \quad : \quad  {1\over \bar\eta^2} \{
{\bar{X}_s} [\tilde\Gamma^+_{2,2} \bar{E}^{(2)+}_{8,t} +
\tilde\Gamma^-_{2,2} \bar{E}^{(2)-}_{8,t}] + {\bar{X}_c}
[\tilde\Gamma^+_{2,2} \bar{E}^{(2)-}_{8,t} + \tilde\Gamma^-_{2,2}
\bar{E}^{(2)+}_{8,t}] \}\ee where \be \tilde\Gamma^{\pm}_{2,2} =
{\tilde\Lambda^+_{2,2} \pm \tilde\Lambda^-_{2,2}\over 2
\bar\eta^2} \ee and \be \bar{E}^{(2)\pm}_{8,t} = {1\over 2}
[\bar{E}_8(\sqrt{q})\pm \bar{E}_8(-\sqrt{q})] \ee

\subsection{Pair Production of small BH's}

Tree-level scattering amplitudes only involving states in the
untwisted sector of the FHSV model are identical to those of the
parent theory with $\cN = 4$ susy. The two-derivative effective
action is expected to receive no quantum corrections, thanks to
$n_v = n_h$, \emph{i.e} the number of vector and hyper multiplets
are always equal, everywhere in the moduli space, including points
of enhanced symmetry \cite{Ferrara:1995yx}. However, twisted
states contribute to higher derivative corrections to the
effective action and can be pair produced at tree-level.

Let us consider the production of two small BPS BH's in the
twisted sector. For scalar states, corresponding to spherically
symmetry small BH's, vertex operators are of the form \be
V^{r,u}_{\Phi, f} = e^{-\varphi} \sigma_f S^r e^{i {\bf
\tilde{p}}_L {\bf Z}_L} :\cB_{N_R^Y}\bar\sigma_f: :\cB_{N_R^Z}
e^{i {\bf \tilde{p}}_R {\bf Z}_R}: :\cB_{N_R^J} \bar\Psi^u:
e^{ipX}\ee where $\sigma_f$ are $Z_2$ twisted fields located at
the fixed point $f$, $S^r$ is an internal $SO(4)$ spin field of
positive chirality, $\bar\Psi^u$ is a primary field of
$E_8^{(2)}$, whose currents are twisted (\ie half-integer modes).
$\cB_{N_R^Y}, \cB_{N_R^Z}, \cB_{N_R^J}$ are polynomials in the
derivatives of the currents $J$ and of the internal coordinates
$Y$ , for $T^4$, and $Z$ for $T^2$. Level matching requires \be
{{1\over 2}}|{\bf\tilde{p}}_L|^2 = {{1\over 2}}
|{\bf\tilde{p}}_R|^2 + {3\over 4} + {N}_R^Z + {N}_R^X + {N}_R^J +
{1\over 4}|{\bf r}|^2 \ee with \be {N}_R^{tot} = {N}_R^Z + {N}_R^X
+ {N}_R^J = \sum_{k=1}^\infty [k n_k^Z + (k - {1\over 2}) (n_{k-
{1\over 2}}^X + n_{k- {1\over 2}}^J)] \ ,\ee that amounts to \be
({\bf m} + {\bf a})({\bf n}+ {\bf b}) = {N}_R^{tot} + {1\over
4}|{\bf r}|^2 + {3\over 4} \ee with $2{\bf a}{\bf b} = 1/2$ (mod
1).

The amplitudes for the processes under consideration
\begin{eqnarray}
\cA=\langle V_{\Phi, f}V_{v/h}V_{v/h} V_{\Phi, f}\rangle
\end{eqnarray}
 can be decomposed into various parts.

In addition to the ubiquitous L-moving contribution $\cE_L(z_i,
p_i)$ that combines with
 \be \langle e^{-\varphi} S^r \sigma_f e^{i {\bf \tilde{p}}_L
{\bf Z}_L}(z_1) e^{-\varphi} S^s \sigma_{f'} e^{-i {\bf \tilde{p}}_L
{\bf Z}_L}(z_4)\rangle = \varepsilon^{rs} \delta_{ff'} z_{14}^{-2-
{{}}|{\bf \tilde{p}}_L|^2} \ee one needs the contribution of the
R-movers
\bea &&\cE_R^{FHSV}(\bar{z}_i, \mathbb{P}_i) = \langle
\cB_{N_R^Y}\bar\sigma_f \cB_{N_R^Z} e^{i {\bf \tilde{p}}_R {\bf
Z}_R} \cB_{N_R^J} \bar\Psi^u
e^{ip_1X_R}(\bar{z}_1)\bar\de\mathbb{X}^M
e^{ik_2X_R}(\bar{z}_2) \nn \\
&&\qquad \bar\de\mathbb{X}^N e^{ik_3X_R}(\bar{z}_3)
\cB_{N_R^Y}\bar\sigma_{f'} \cB_{N_R^Z} e^{-i {\bf \tilde{p}}_R
{\bf Z}_R}\cB_{N_R^J}
\bar\Psi^\dagger_ve^{ip_4X_R}(\bar{z}_4)\rangle \eea

Let us consider for for simplicity the case of
small BH's charged wrt the `visible' gauge group $E_8$, whose
vertex operator involves a primary field $\bar{\Psi}^u$ ($N_R^J = 0$) of
dimension \be h_{\Psi} = {1\over 2} +  {1\over 4}|{\bf r}|^2 \ee
in a representation $\bf R$, with highest weight $|\bf r|$, of
 the $E_8$ current algebra at level $\kappa =2$. For gauge
bosons in the initial state, $\bar\de\mathbb{X}^{M/N} \rightarrow
\bar{J}^{a/b}$, one needs the correlation function \bea
&&\bar\cC_R^{ab,u}{}_v (\bar{z}_i, p_i)= \langle \bar\Psi^u
(\bar{z}_1)\bar{J}^a(\bar{z}_2) \bar{J}^b (\bar{z}_3)
\bar\Psi^\dagger_v (\bar{z}_4) \rangle \langle
e^{ip_1X_R}(\bar{z}_1) e^{ik_2X_R}(\bar{z}_2)
e^{ik_3X_R}(\bar{z}_3)
e^{ip_4X_R}(\bar{z}_4)\rangle = \nn \\
&& {N_\Psi \over \bar{z}_{23}^2 \bar{z}_{14}^{2h_{\Psi}}} \left[
2\delta^{ab} \delta^{u}{}_v + [t^a, t^b]^u{}_v \left(
{\bar{z}_{12} \bar{z}_{34} \over \bar{z}_{13} \bar{z}_{24}} -
{\bar{z}_{13} \bar{z}_{24} \over \bar{z}_{12} \bar{z}_{34}}\right)
+ \{t^a, t^b\}^u{}_v \left( {\bar{z}_{12} \bar{z}_{34} \over
\bar{z}_{13} \bar{z}_{24}} + {\bar{z}_{13} \bar{z}_{24} \over
\bar{z}_{12} \bar{z}_{34}}\right) \right] \cI_R(\bar{z}_i,
p_i)\nn\\ \eea where $N_\Psi$ is some normalization and $\cI_R$ is
the R-mover Koba-Nielsen factor.

For gravitons in the initial state, $\bar\de\mathbb{X}^{M/N}
\rightarrow \bar\de X^{\mu/\nu}$, one needs instead \bea
&&\bar\cG_R^{\mu\nu,u}{}_v (\bar{z}_i, p_i) = \langle \bar\Psi^u
(\bar{z}_1) \bar\Psi^\dagger_v (\bar{z}_4) \rangle \langle
e^{ip_1X_R}(\bar{z}_1)\bar\de X^\mu e^{ik_2X_R}(\bar{z}_2)
\bar\de X^\nu e^{ik_3X_R}(\bar{z}_3) e^{ip_4X_R}(\bar{z}_4)\rangle \nn \\
&&= - { N_\Psi \delta^{u}{}_v\over \bar{z}_{14}^{2h_{\Psi}}}
\left[{\eta^{\mu\nu}\over \bar{z}^2_{23}}  + \sum_{i\neq 2}{
p_i^\mu\over \bar{z}_{2i}}\sum_{j\neq 3}{ p_j^\mu\over
\bar{z}_{3j}}\right]\cI_R(\bar{z}_i, p_i)
 \eea

The remaining correlation function factorizes and yields \bea
&&\cK_R(\bar{z}_i, {\bf \tilde{p}}_R) = \langle
\cB_{N_R^Y}\bar\sigma_f\cB_{N_R^Z} e^{i {\bf \tilde{p}}_R {\bf
Z}_R}(\bar{z}_1)\cB_{N_R^Y}\bar\sigma_{f'}
 \cB_{N_R^Z} e^{-i {\bf \tilde{p}}_R
{\bf Z}_R}(\bar{z}_4)\rangle = {N_{YZ}({\bf \tilde{p}}_R)\over
\bar{z}_{14}^{2N_R^Y + {1\over 2} + 2N_R^Z +|{\bf \tilde{p}}_R|^2} }\nn \\
\eea Combining with the L-mover contribution, the computation
proceeds as in the toroidal case.

For small BH's in the twisted sector of the FHSV model, the
amplitude for pair production in graviton scattering is the same
as in the untwisted sector, up to some (moduli dependent)
normalization.

As in toroidal compactifications, pair production of small BH's in
gauge boson scattering is very sensitive to the `gauge' quantum
numbers of the products. For the above simple case in which the
charged BH's correspond to primary fields of the current algebra
the resulting scattering amplitude is \be \cA = g_s^2 (2\pi)^{4}
\delta^4(\Sigma_i p_i) {{{}}} \delta_{f_1f_4}\epsilon^{r_1r_4}
N({\bf{\tilde{p}}}_R) [\delta^{ab}\delta^u{}_v A_0 + (t^a
t^b)^u{}_v A_1 + (t^b t^a)^u{}_v A_{-1} ] \ee where $A_n(p_i,
a_i)$ are defined in Eq (\ref{An}). Notice that $A_1$ and $A_{-1}$
get exchanged under $2 \leftrightarrow 3$ or, equivalently, $1
\leftrightarrow 4$.

For generic small BH with the same charges (${\bf m}, {\bf n},
{\bf r}$) and mass, level matching allows for descendants under
the action of the current algebra. We expect this to change at
most the normalization constant. After summing over all the
degenerate `final' states, one finds a large multiplicity due to
the microscopic entropy of the small BH states.

\section{Conclusions and oulook}
\label{Conc}

Let us conclude with few comments and directions for future
investigation.

Our results for pair production of small BH's with two charges,
corresponding to perturbative BPS states in toroidal
compactifications of heterotic strings or in the FHSV model,
display a certain degree of universality. The presence of Regge
poles and soft UV behavior are a hallmark of any string amplitude.
In more realistic string models describing \eg collisions at LHC,
one should convolute `partonic' cross sections for BH production,
such as the ones computed here, with the parton distributions
inside the protons, that may significantly change the shape of the
signal. The results also depend on the particles initiating or
mediating the process, gravitons or vector bosons. Not
surprisingly, for gravitons we found a huge suppression related to
the smallness of Newton's constant. In perturbative heterotic
strings, scenari with LED are hard to accommodate. Anyway the
relative suppression wrt processes mediated by `visible' gauge
bosons is $1/M_s^2$. Needless to say we don't find any increase of
the cross section with the CM energy as in FT toy models. Although
we focussed on the lowest spin scalar components of BPS BH
multiplets, our analysis applies to the spin 1/2 and 1
superpartners. In any case the corresponding BH solutions have
non-rotating horizons, since the broken supersymmetry parameters
generating BPS BH supermultiplets vanish at the
horizon\footnote{We thank Ashoke Sen for a clarifying discussion
on this and related issues.}

We expect the same qualitative behavior for the production of
extremal non BPS BH's with $N_R=1$ and $M^2 = |{\bf p}_R|^2 =
|{\bf p}_L|^2 + N_L - \delta_L$, whereby the Left movers are not
in their ground states. The world-sheet computations would be very
similar and reliable. Extremal BH's -- be BPS or not -- cannot
decay since there are no states with the same charges and lower
mass. They do not emit Hawking radiation. In some cases they admit
a fuzzball description \cite{Mathur:2008nj, Skenderis:2008qn}.

A quantitatively different but qualitatively similar story applies
to highly excited string states which are far from the BPS bound
\cite{Chialva:2004xm, Iengo:2006gm}. While, it is believed anyway
that the string/BH correspondence principle should work even in
this case \cite{Veneziano:2008xa, Veneziano:2008zb}, far from
extremal BH's have properties, including their mass and entropy,
that are sensitive to the string coupling and other freely
adjustable parameters (moduli fields). Although the grey-body
factor for near extremal BH's has been very successfully computed
more than ten years ago \cite{Maldacena:1997tm}, dynamical
properties of (small) non-extremal BH's corresponding to
fundamental heterotic strings have been only explored in the
recent past. In particular in \cite{Cornalba:2006hc}, convincing
evidence was given in favour of a string/BH transition in
dynamical processes of emission and absorption. Furthermore, a
dynamical analysis, similar to the one performed here, elucidated
the distribution in size and typical configuration of very massive
closed string states as a function of the string coupling
\cite{Chialva:2009pf, Chialva:2009pg}.

Finally, we would like to briefly comment on non spherically
symmetric \eg rotating BH's. It is well known that string
excitations at level $N$ can carry high spin. The maximal spin is
$J_{Max} \approx N$. BPS states with higher spin are described by
vertex operators of the form \be V_{\cH}^s = \cH_{i,\mu_1 ...
\mu_s} e^{-\varphi} \psi^i e^{i{\bf p}_L {\bf X}_L} \bar\de
X^{\mu_1} ... \bar\de X^{\mu_s} e^{i{\bf p}_R {\bf X}_R} e^{i p
X}\ee The tensor $\cH_{i,\mu_1... \mu_s}$ is totally symmetric by
construction in the $\mu$ indices and, in order for the state to
be BRS invariant, it should be transverse $p^\mu \cH_{i,\mu \mu_2
... \mu_s} = 0$ and traceless. It is well possible that BPS
strings with high angular momentum be in correspondence with black
rings rather than BH's \cite{Elvang:2004rt, Bena:2004de,
Dabholkar:2006za, Bena:2007kg}. Moreover the validity of the
string/BH correspondence principle may need to be reconsidered for
states with angular momentum since varying the moduli states with
high spin can `decay' or rather transform into bound states of
components with lower spin.

 \section*{Acknowledgments}
Fruitful discussions with S.~Ferrara, F.~Fucito, R.~Kallosh,
A.~Lionetto, J.~F.~Morales, D.~Ricci Pacifici, A.~V.~Santini,
A.~Sen are kindly acknowledged. This work was partially supported
by the ERC Advanced Grant n.226455 {\it ``Superfields''} and by
the Italian MIUR-PRIN contract 2007-5ATT78 {\it ``Symmetries of
the Universe and of the Fundamental Interactions''}.

\end{document}